\pdfoutput=1
\newcommand*{\ATLASLATEXPATH}{latex/}
\documentclass[UKenglish,texlive=2016,txfonts,cernpreprint]{\ATLASLATEXPATH atlasdoc}

\usepackage{\ATLASLATEXPATH atlaspackage}
\usepackage{\ATLASLATEXPATH atlasbiblatex}

\usepackage{\ATLASLATEXPATH atlascontribute}

\usepackage[BSM]{\ATLASLATEXPATH atlasphysics}

\graphicspath{{logos/}{.}}

\usepackage{paper-defs}

\usepackage{enumitem}

\newcommand{\HistFitter}{{\textsc{HistFitter}}\xspace}

\AtlasTitle{\boldmath Search for supersymmetry in events with four or more leptons in $\sqrt{s}=13\TeV$ $pp$ collisions with ATLAS}

\author{The ATLAS Collaboration}

\usepackage{authblk}
\AtlasRefCode{SUSY-2016-21}

\PreprintIdNumber{CERN-EP-2017-300}

\AtlasJournalRef{\PRD ~98 (2018) 032009}
\AtlasDOI{10.1103/PhysRevD.98.032009}

\AtlasAbstract{%
Results from a search for supersymmetry in events with four or more charged leptons (electrons, muons and taus) are presented. 
The analysis uses a data sample corresponding to $36.1\,\ifb$  of proton--proton collisions delivered by the Large Hadron Collider at $\sqrt{s}=13\TeV$ and recorded by the ATLAS detector.
Four-lepton signal regions with up to two hadronically decaying taus are designed to target a range of supersymmetric scenarios that can be either enriched in or depleted of events involving the production and decay of a $Z$ boson. 
Data yields are consistent with Standard Model expectations and results are used to set upper limits on the event yields from processes beyond the Standard Model. 
Exclusion limits are set at the 95\% confidence level in simplified models of General Gauge Mediated supersymmetry, where higgsino masses are excluded up to $295\GeV$.
In $R$-parity-violating simplified models with decays of the lightest supersymmetric particle to charged leptons, lower limits of 
$1.46\TeV$, 
$1.06\TeV$, and 
$2.25\TeV$ 
are placed on wino, slepton and gluino masses, respectively. 
}

\usepackage{adjustbox}
\usepackage{arrayjobx}

\usepackage{siunitx}
\hypersetup{pdftitle={ATLAS draft},pdfauthor={The ATLAS Collaboration}}

\usepackage{longtable}
\usepackage{rotating}
\usepackage{makecell}

\begin{document}

\maketitle


\section{Introduction}\label{sec:intro}

Supersymmetry (SUSY)~\cite{Golfand:1971iw,Volkov:1973ix,Wess:1974tw,Wess:1974jb,Ferrara:1974pu,Salam:1974ig} is a space-time symmetry that postulates the existence of new particles with spin differing by one half-unit from their Standard Model (SM) partners. 
In supersymmetric extensions of the SM, each SM fermion (boson) is associated with a SUSY boson (fermion), having the same quantum numbers as its partner except for spin.
The introduction of these new SUSY particles provides a potential solution to the hierarchy problem~\cite{Sakai:1981gr,Dimopoulos:1981yj,Ibanez:1981yh,Dimopoulos:1981zb}.

The scalar superpartners of the SM fermions are called sfermions (comprising the charged sleptons, $\slepton$, the sneutrinos, $\snu$, and the squarks, $\squark$), while the gluons have fermionic superpartners called gluinos ($\gluino$). 
The bino, wino and higgsino fields are fermionic superpartners of the SU(2)$\times$U(1) gauge fields of the SM, and the two complex scalar doublets of a minimally extended Higgs sector, respectively. 
Their mass eigenstates are referred to as charginos $\tilde{\chi}_{i}^{\pm}$ $(i=1,2)$ and neutralinos $\tilde{\chi}_{j}^{0}$ $(j=1,2,3,4)$, numbered in order of increasing mass. 

In the absence of a protective symmetry, SUSY processes not conserving lepton number ($L$) and baryon number ($B$) could result in proton decay at a rate that is in conflict with the tight experimental constraints on the proton lifetime~\cite{Ahmed:2003sy}.
This conflict can be avoided by imposing the conservation of $R$-parity~\cite{Farrar:1978xj}, defined as $(-1)^{3(B-L)+2S}$, where $S$ is spin, 
or by explicitly conserving either $B$ or $L$ in the Lagrangian in $R$-parity-violating (RPV) scenarios.
In RPV models, the lightest SUSY particle (LSP) is unstable and decays to SM particles, 
including charged leptons and neutrinos when violating $L$ but not $B$. 
In $R$-parity-conserving (RPC) models, the LSP is stable and leptons can originate from unstable weakly interacting sparticles decaying into the LSP. 
Both the RPV and RPC SUSY scenarios can therefore result in signatures with high lepton multiplicities and substantial missing transverse momentum, 
selections on which can be used to suppress SM background processes effectively.

This paper presents a search for new physics in final states with at least four isolated, charged leptons (electrons, muons or taus) where up to two hadronically decaying taus are considered. 
The analysis exploits the full proton--proton dataset collected by the ATLAS experiment during the 2015 and 2016 data-taking periods, 
corresponding to an integrated luminosity of $36.1~\ifb$ at a center-of-mass energy of $13\TeV$.
The search itself is optimized using several signal models but is generally model-independent, using selections on the presence or 
absence of $Z$ bosons in the event and loose requirements on effective mass or missing transverse momentum.
Results are presented in terms of the number of events from new physics processes with a four charged lepton signature, and also in terms of RPV and RPC SUSY models.

Previous searches for SUSY particles using signatures with three or more leptons were carried out 
at the Tevatron collider~\cite{Abazov:2005ku,Abazov:2009zi,Abazov:2006nw,Aaltonen:2007an,Aaltonen:2008pv,Abulencia:2007mp},
and at the LHC by the ATLAS experiment~\cite{SUSY-2012-17,SUSY-2011-25,SUSY-2013-12,SUSY-2013-13} 
and the CMS experiment~\cite{CMS-SUS-10-008,CMS-SUS-11-013,CMS-SUS-12-006,CMS-SUS-13-003,CMS-SUS-13-002}.
This analysis closely follows the $7\TeV$~\cite{SUSY-2012-17} and $8\TeV$~\cite{SUSY-2013-13} ATLAS analyses.

\section{SUSY scenarios}\label{sec:susymodels}

SUSY models are used for signal region optimization and to interpret the results of this analysis.
Models of both RPV SUSY and RPC SUSY are considered here, as they each require a different approach for signal selection, as discussed in Section~\ref{sec:eventreco}.

In all scenarios, the light CP-even Higgs boson, $h$, of the Minimal Supersymmetric extension of the SM~\cite{Fayet:1976et,Fayet:1977yc} Higgs sector is assumed to be practically identical to the SM Higgs boson~\cite{Carena:2013ytb}, 
with the same mass and couplings as measured at the LHC~\cite{Aad:2012tfa,Chatrchyan:2012xdj,ATLAS-CONF-2015-044}.
In addition, the decoupling limit is used, which is defined by $m_A \gg m_Z$, while the CP-odd ($A$), the neutral CP-even ($H$), and the two charged ($H^\pm$) Higgs bosons are considered to be very heavy and thus considerably beyond the kinematic reach of the LHC.

\subsection{RPV SUSY scenarios}
\label{sec:rpvsusysignal}

In generic SUSY models with minimal particle content, the superpotential includes terms that violate conservation of $L$ and $B$~\cite{Weinberg:1981wj,Sakai:1981pk}:

\vspace{-0.4cm}
\begin{equation*}\label{eq:rpvpotential}
\frac{1}{2}\lam L_iL_j\bar{E}_k + 
\lam' L_iQ_j\bar{D}_k + 
\frac{1}{2}\lam'' \bar{U_i}\bar{D_j}\bar{D_k} +
\kappa_iL_iH_2,
\end{equation*}

\vspace{-0.2cm}
\noindent
where $L_i$ and $Q_i$ indicate the lepton and quark $\mathrm{SU}(2)$-doublet superfields, respectively, and $\bar{E_i}$, $\bar{U_i}$ and $\bar{D_i}$ are the corresponding singlet superfields. 
Quark and lepton generations are referred to by the indices $i$, $j$ and $k$, while the Higgs field that couples to up-type quarks is represented by the Higgs $\mathrm{SU(2)}$-doublet superfield $H_2$. 
The \lam, $\lam'$ and $\lam''$ parameters are three sets of new Yukawa couplings, while the $\kappa_i$ parameters have dimensions of mass.

Simplified models of RPV scenarios are considered, where the LSP is a bino-like neutralino ($\ninoone$) and decays via an RPV interaction.
The LSP decay is mediated by the following lepton-number-violating superpotential term:

\vspace{-0.8cm}
\begin{equation*}
\label{eq:LLERPV}
W_{\LLE} = \frac{1}{2}\lam L_iL_j\bar{E}_k.
\end{equation*}

\vspace{-0.2cm}
This RPV interaction allows the following decay of the neutralino LSP:

\vspace{-0.6cm}
\begin{equation}
\label{eq:LLERPV_N1decay}
\ninoone \to \ell^{\pm}_k \ell^{\mp}_{i/j} \nu_{j/i},
\end{equation}

\vspace{-0.2cm}
through a virtual slepton or sneutrino, with the allowed lepton flavors depending on the indices of the associated \lam couplings~\cite{Barbier:2004ez}.
The complex conjugate of the decay in Eq.~\eqref{eq:LLERPV_N1decay} is also allowed. 
Thus, in the case of pair production, every signal event contains a minimum of four charged leptons and two neutrinos. 

In principle, the nine\footnote{reduced from 27 by the antisymmetry requirement $\lambda_{ijk} = -\lambda_{jik}$} \lam RPV couplings allow the \ninoone to decay to every possible combination of charged lepton pairs, where the branching ratio for each combination differs for each \lam.
For example, for $\lambda_{121}\neq 0$ the branching ratios for $\ninoone \ra e\mu\nu$, $\ninoone \ra ee\nu$ and $\ninoone \ra \mu\mu\nu$ are 50\%, 50\% and 0\% respectively,
whereas for $\lambda_{122}\neq 0$ the corresponding branching ratios are 50\%, 0\% and 50\%.
In Ref.~\cite{SUSY-2013-13}, it was found that the four-charged-lepton search sensitivity is comparable in the cases of $\lambda_{121}\neq 0$ or $\lambda_{122}\neq 0$, and for $\lambda_{133}\neq 0$ or $\lambda_{233}\neq 0$.
Since the analysis reported here uses similar techniques, the number of $L$-violating RPV scenarios studied is reduced by making no distinction between the electron and muon decay modes of the $\ninoone$.
Two extremes of the \lam RPV couplings are considered: 
\begin{itemize}
\item $\LLE 12k$ ($k \in {1,2}$) scenarios, where $\lambda_{12k}\neq 0$ and only decays to electrons and muons are included,
\item $\LLE i33$ ($i \in {1,2}$) scenarios, where $\lambda_{i33}\neq 0$ and only decays to taus and either electrons or muons are included.
\end{itemize}
In both cases, all other RPV couplings are assumed to be zero. 
The branching ratios for the $\ninoone$ decay in the $\LLE 12k$ and $\LLE i33$ are shown in \Tab{\ref{tab:datasets_RPVsignalBR}}.
The sensitivity to $\lambda$ couplings not considered here (e.g.\ $\lambda_{123}$) is expected to be between that achieved in the $\LLE 12k$ and $\LLE i33$ scenarios.

\begin{table}[ht]
\centering
\begin{tabular}{l c c c}
\toprule
Scenario & \multicolumn{3}{c}{\ninoone branching ratios} \\
\midrule
$\LLE 12k$ & $  \ee\nu$ (1/4) & $  \emu\nu$ (1/2) & $ \mumu\nu$ (1/4) \\
$\LLE i33$ & $  \etau\nu$ (1/4) & $\tautau\nu$ (1/2) & $\mutau\nu$ (1/4) \\
\bottomrule
\end{tabular}
\caption{Decay modes and branching ratios for the \ninoone LSP in the RPV models, where $\nu$ denotes neutrinos or antineutrinos of any lepton generation. 
\label{tab:datasets_RPVsignalBR}}
\end{table}

For the pure-bino $\ninoone$ considered here, the $\ninoone\ninoone$ production cross section is found to be vanishingly small, thus models that include one or more next-to-lightest SUSY particles (NLSP) 
are considered in order to obtain a reasonably large cross section. 
The choice of NLSP in the $\LLE 12k$ and $\LLE i33$ scenarios determines the cross section of the SUSY scenario, and can impact the signal acceptance to a lesser extent.
In all cases, the NLSP is pair-produced in an RPC interaction, and decays to the LSP (which itself undergoes an RPV decay).
Three different possibilities are considered for the NLSP in the $\LLE 12k$ and $\LLE i33$ scenarios:
\begin{itemize}
  \item \textbf{Wino NLSP}: mass-degenerate wino-like charginos and neutralinos are produced in association ($\chinoonep\chinoonem$ or $\chinoonepm\ninotwo$). 
   The \chinoonepm (\ninotwo) decays to the LSP while emitting a \Wboson (\Zboson or \Higgs) boson, as shown in Figures~\ref{fig:RPVDiagrams_winoZ} and \ref{fig:RPVDiagrams_winoH}.
  \item $\pmb{\sleptonL}/\pmb{\snu}$~\textbf{NLSP}: mass-degenerate left-handed sleptons and sneutrinos of all three generations 
  are produced in association ($\sleptonL\sleptonL$, $\snu\snu$, $\sleptonL\snu$). 
  The \sleptonL (\snu) decays to the LSP while emitting a charged lepton (neutrino) as seen in Figure~\ref{fig:RPVDiagrams_LHSlepSnu}.
  \item $\mathbf{\gluino}$~\textbf{NLSP}:   gluino pair-production, where the gluino decays to the LSP while emitting a quark--antiquark pair ($u,d,s,c,b$ only, 
  	with equal branching ratios), as seen in Figure~\ref{fig:RPVDiagrams_gluino}.
  \end{itemize}
For the RPV models, the LSP mass is restricted to the range $10\GeV \leq m(\text{LSP}) \leq m(\text{NLSP}) - 10\GeV$ to ensure that both the RPC cascade decay and the RPV LSP decay are prompt.
Non-prompt decays of the $\ninoone$ in similar models were previously studied in Ref.~\cite{SUSY-2014-02}.

\begin{figure}[htbp]
\begin{center}
\subfloat[][Wino \Wboson/\Zboson NLSP]{\includegraphics[width=0.35\textwidth]{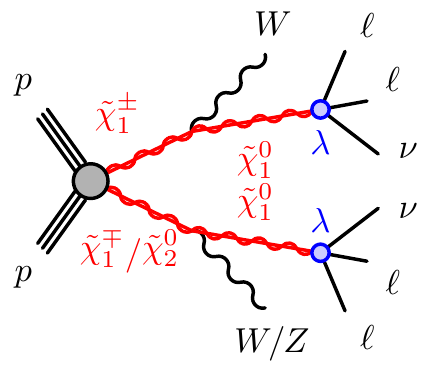}\label{fig:RPVDiagrams_winoZ}}\hspace{0.1\textwidth}
\subfloat[][Wino \Wboson/\Higgs NLSP]{\includegraphics[width=0.35\textwidth]{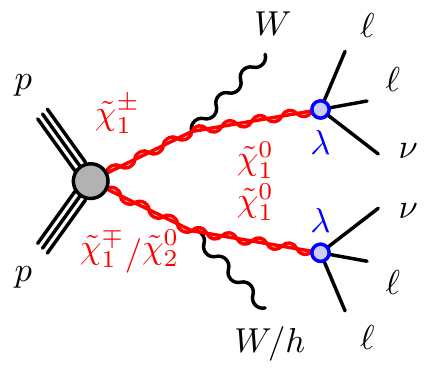}\label{fig:RPVDiagrams_winoH}}\\
\subfloat[][$\sleptonL/\snu$ NLSP]{\includegraphics[width=0.35\textwidth]{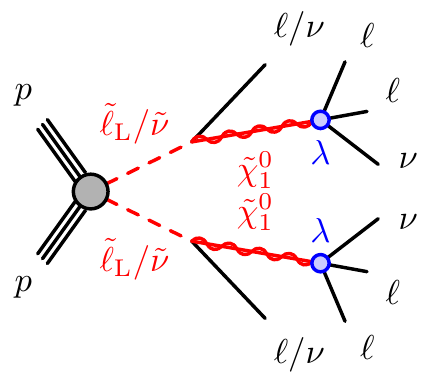}\label{fig:RPVDiagrams_LHSlepSnu}}\hspace{0.1\textwidth}
\subfloat[][$\gluino$ NLSP]{\includegraphics[width=0.35\textwidth]{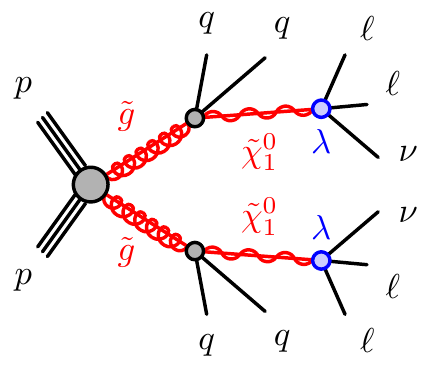}\label{fig:RPVDiagrams_gluino}}
\end{center}
\caption{Diagrams of the benchmark SUSY models of RPC NLSP pair-production of 
\protect\subref{fig:RPVDiagrams_winoZ}-\protect\subref{fig:RPVDiagrams_winoH}~a wino, 
\protect\subref{fig:RPVDiagrams_LHSlepSnu}~slepton/sneutrino and
\protect\subref{fig:RPVDiagrams_gluino}~gluino, followed by the RPV decay of the $\ninoone$ LSP. 
The LSP is assumed to decay as $\ninoone\ra \ell\ell\nu$ with 100\% branching ratio. }
\label{fig:RPVDiagrams}
\end{figure}

\subsection{RPC SUSY scenarios}
\label{sec:rpcsusysignal}

RPC scenarios with light $\ninoone$, $\ninotwo$ and $\chinoonepm$ higgsino states are well motivated by naturalness~\cite{Barbieri:1987fn,deCarlos:1993yy}.
However, they can be experimentally challenging, as members of the higgsino triplet are close in mass and decays of the $\ninotwo/\chinoonepm$ to a $\ninoone$ LSP result in low-momentum decay products that are difficult to reconstruct efficiently. 
Searches for higgsino-like $\chinoonepm$ in approximately mass-degenerate scenarios were performed by the LEP experiments, where chargino masses below $103.5\GeV$ were excluded~\cite{LEPSUSYWG:01-03.1} (reduced to $92\GeV$ for chargino--LSP mass differences between $0.1\GeV$ and $3\GeV$).
Recently, the ATLAS experiment has excluded higgsino-like $\ninotwo$ up to masses $\sim$$145\GeV$ and down to $\ninotwo$--LSP mass differences of $2.5\GeV$~\cite{Aaboud:2017leg} for scenarios where the $\chinoonepm$ mass is assumed to be half-way between the two lightest neutralino masses. 
In the Planck-scale-mediated SUSY breaking scenario the gravitino $\gravino$ is the fermionic superpartner of the graviton, 
and its mass is comparable to the masses of the other SUSY particles, $m \sim 100\GeV$~\cite{Lahanas:1986uc,NILLES19841}. 
General Gauge Mediated (GGM) SUSY models~\cite{Meade:2008wd} predict the $\gravino$ is nearly massless and offer an opportunity to study light higgsinos.
The decays of the higgsinos to the LSP $\gravino$ would lead to on-shell $Z/h$, and the decay products can be reconstructed.

Simplified RPC models inspired by GGM are considered here, where the only SUSY particles within reach of the LHC are an almost mass-degenerate higgsino triplet $\chinoonepm,\,\ninoone,\,\ninotwo$ and a massless $\gravino$.
To ensure the SUSY decays are prompt, the $\chinoonepm$ and $\ninotwo$ masses are set to $1\GeV$ above the $\ninoone$ mass, and due to their weak coupling with the gravitino always decay to the $\ninoone$ via virtual $Z/W$ bosons (which in turn decay to very soft final states that are not reconstructed). 
The $\ninoone$ decays promptly to a gravitino plus a $Z$ or $h$ boson, $\ninoone \ra Z/h + \gravino$, where the leptonic decays of the $Z/h$ are targeted in this analysis. 
Four production processes are included in this higgsino GGM model: $\chinoonep\chinoonem$, $\chinoonepm\ninoone$, $\chinoonepm\ninotwo$ and $\ninoone\ninotwo$, as shown in Figure~\ref{fig:HinoGGMdiagrams}, and the total SUSY cross section is dominated by $\chinoonepm\ninoone$ and $\chinoonepm\ninotwo$ production.
The $\ninoone \ra Z \gravino$ branching ratio is a free parameter of the GGM higgsino scenarios, and so offers an opportunity to study 4$\ell$ signatures with one or more $Z$ candidates. 

\begin{figure}[htbp]
\begin{center}
\subfloat[][]{\includegraphics[width=0.4\textwidth]{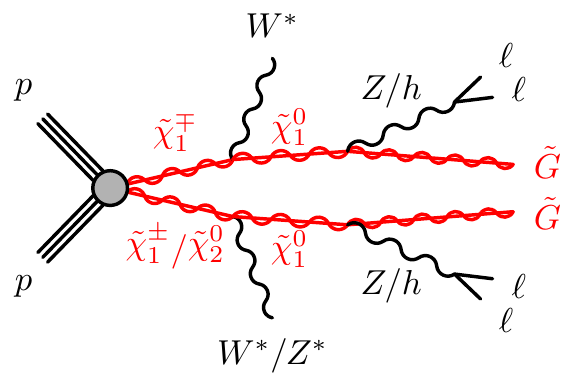}}
\subfloat[][]{\includegraphics[width=0.4\textwidth]{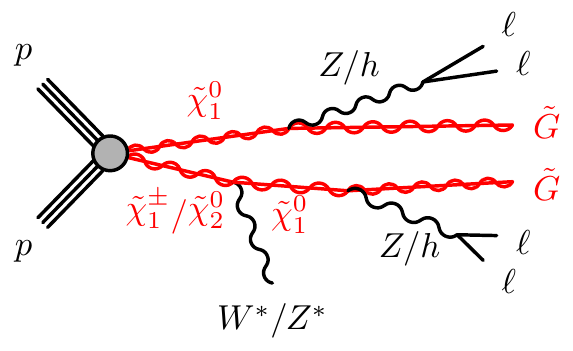}}
\end{center}
\caption{Diagrams of the processes in the SUSY RPC GGM higgsino models. 
The $W^{*}/\,Z^{*}$ produced in the $\chinoonepm/\ninotwo$ decays are off-shell ($m\sim 1\GeV$) and their decay products are usually not reconstructed.}
\label{fig:HinoGGMdiagrams}
\end{figure}

\section{The ATLAS detector}\label{sec:atlas}
The ATLAS detector~\cite{PERF-2007-01} is a multipurpose particle physics detector with forward-backward symmetric cylindrical geometry.\footnote{ATLAS uses a right-handed coordinate system with its origin at the nominal interaction point (IP) 
in the center of the detector and the $z$-axis along the beam pipe. 
The $x$-axis points from the IP to the center of the LHC ring, and the $y$-axis points upward. 
Cylindrical coordinates $(r,\phi)$ are used in the transverse plane, $\phi$ being the azimuthal angle around the $z$-axis. 
The pseudorapidity is defined in terms of the polar angle $\theta$ as $\eta=-\ln\tan(\theta/2)$.
Rapidity is defined as $y=0.5 \ln\left[(E+p_z)/(E-p_z)\right]$ where $E$ denotes the energy and $p_z$ is the component of the momentum along the beam direction.
}
 The inner tracking detector (ID) covers $|\eta|<2.5$ and consists of a silicon pixel detector, a semiconductor microstrip detector, and a transition radiation tracker. 
The innermost pixel layer, the insertable B-layer~\cite{Blayer}, was added for the $\sqrt{s}=13\TeV$ running period of the LHC.
The ID is surrounded by a thin superconducting solenoid providing a 2$\,$T axial magnetic field. A high-granularity lead/liquid-argon sampling calorimeter measures the energy and the position of electromagnetic showers within $|\eta|<3.2$. 
Sampling calorimeters with liquid argon as the active medium are also used to measure hadronic showers in the endcap ($1.5<|\eta|<3.2$) and forward ($3.1<|\eta|<4.9$) regions, while a steel/scintillator tile calorimeter measures hadronic showers in the central region ($|\eta|<1.7$). 
The muon spectrometer (MS) surrounds the calorimeters and consists of three large superconducting air-core toroid magnets, each with eight coils, a system of precision tracking chambers ($|\eta|<2.7$), and fast trigger chambers ($|\eta|<2.4$). 
A two-level trigger system~\cite{ATL-DAQ-PUB-2016-001} selects events to be recorded for offline analysis.

\section{Monte Carlo simulation}\label{sec:mcsim}

Monte Carlo (MC) generators were used to simulate SM processes and new physics signals. 
The SM processes considered are those that can lead to signatures with at least four reconstructed charged leptons. 
Details of the signal and background MC simulation samples used in this analysis, as well as the order of cross section calculations in perturbative QCD used for yield normalization, are shown in Table~\ref{tab:SMbkgMC}.
Signal cross sections were calculated to next-to-leading order in the strong coupling constant, adding the resummation of soft gluon emission at next-to-leading-logarithmic accuracy (NLO+NLL)~\cite{Beenakker:1996ch,Kulesza:2008jb,Kulesza:2009kq,Beenakker:2009ha,Beenakker:2011fu,Fuks:2012qx,Fuks:2013vua,Fuks:2013lya}. 
The nominal signal cross section and its uncertainty were taken from an envelope of cross section predictions using different parton distribution function (PDF) sets and factorization and renormalization scales, as described in Ref.~\cite{Borschensky:2014cia}.

The dominant irreducible background processes that can produce four prompt and isolated charged leptons are $ZZ$, $\ttbar Z$, $VVV$ and Higgs production (where $V=W,\,Z$, and includes off-shell contributions). 
For the simulated $ZZ$ production, the matrix elements contain all diagrams with four electroweak vertices, and they were calculated for up to one extra parton at NLO, and up to three extra partons at LO. 
The production of top quark pairs with an additional $Z$ boson was simulated with the cross section normalized to NLO. 
Simulated triboson ($VVV$) production includes the processes $ZZZ$, $WZZ$ and $WWZ$ with four to six charged leptons, and was generated at NLO with additional LO matrix elements for up to two extra partons. 
The simulation of Higgs processes includes Higgs production via gluon--gluon fusion ($ggH$) and vector-boson fusion (VBF), and associated production with a boson ($WH$, $ZH$) or a top--antitop pair ($\ttbar H$). 
Other irreducible background processes with small cross sections are grouped into a category labeled ``Other'', which contains the \tWZ, \ttWW, $tttW$ and \tttt processes.

For all MC simulation samples, the propagation of particles through the ATLAS detector was modeled with \GEANT4~\cite{Agostinelli:2002hh} using the full ATLAS detector simulation~\cite{SOFT-2010-01}, or a fast simulation using a parameterization of the response of the electromagnetic and hadronic calorimeters~\cite{SOFT-2010-01} and \GEANT4 elsewhere. 
The effect of multiple proton--proton collisions in the same or nearby bunch crossings, in-time and out-of-time pileup, 
is incorporated into the simulation by overlaying additional minimum-bias events generated with \PYTHIA8~\cite{Sjostrand:2014zea} onto hard-scatter events.
Simulated events are reconstructed in the same manner as data, and are weighted to match the distribution of the expected mean number of interactions per bunch crossing in data.
The simulated MC samples are corrected to account for differences from the data in the triggering efficiencies, 
lepton reconstruction efficiencies, and the energy and momentum measurements of leptons and jets. 

\begin{table}[h]
\centering
\scriptsize
\begin{tabular}{ l|ccccc }
	\toprule
	Process                      &                               Generator(s)                               & Simulation &                                                             Cross-section                                                             &               Tune               &             PDF set             \\
	                             &                                                                          &                &                                                              calculation                                                              &                                  &  \\ \midrule
	\WZ, \WW                     &                  \SHERPA 2.2.1~\cite{Gleisberg:2008ta}                   &    Full     &                                                   NLO~\cite{ATL-PHYS-PUB-2016-002}                                                    &            \SHERPA default              & NNPDF30NNLO~\cite{Ball:2014uwa} \\
\ZZ                        &                  \SHERPA 2.2.2~\cite{Gleisberg:2008ta}                   &    Full     &                                                   NLO~\cite{ATL-PHYS-PUB-2016-002}                                                    &             \SHERPA default              & NNPDF30NNLO~\cite{Ball:2014uwa} \\ \midrule
	\VVV                         &                              \SHERPA 2.2.1                               &    Full     &                                                   NLO~\cite{ATL-PHYS-PUB-2016-002}                                                    &             \SHERPA default              &           NNPDF30NNLO           \\ \midrule
	$ggH$, VBF, $ggZH$           & \POWHEG v2~\cite{Alioli:2010xd} + \PYTHIA 8.186~\cite{Sjostrand:2014zea} &    Full     &                                                  NNLO+NNLL~\cite{deFlorian:2016spz}                                                   &    AZNLO~\cite{STDM-2012-23}     &     CT10~\cite{Lai:2010vv}      \\
	$ZH$, $WH$                   &                              \PYTHIA 8.186                               &    Full     &                                                               NNLO+NNLL                                                               & A14~\cite{ATL-PHYS-PUB-2014-021} &            NNPDF23LO            \\
	$\ttbar H$                   &                 \MGMCatNLO 2.3.2~\cite{Frixione:2002ik}                  &    Full     &                                                   NLO~\cite{ATL-PHYS-PUB-2016-005}                                                    &               A14                &  NNPDF23LO~\cite{Ball:2012cx}   \\
	                             &                             + \PYTHIA 8.186                              &                &                                                                                                                                       &                                  &  \\ \midrule
	\ttZ, \ttW, \ttWW            &                  \MGMCatNLO~2.2.2~\cite{Alwall:2014hca}                  &    Full     &                                                   NLO~\cite{ATL-PHYS-PUB-2016-005}                                                    &               A14                &            NNPDF23LO            \\
	                             &                             + \PYTHIA 8.186                              &                &                                                                                                                                       &                                  &  \\
	\ttZ{$^\mathbf{\dagger}$}    &                              \SHERPA 2.2.1                               &     Fast      &                                                   NLO~\cite{ATL-PHYS-PUB-2016-005}                                                    &             \SHERPA default              &           NNPDF30NNLO           \\ \midrule
	\tWZ                         &                     \AMCatNLO 2.3.2 + \PYTHIA 8.186                      &    Full     &                                                   NLO~\cite{ATL-PHYS-PUB-2016-005}                                                    &               A14                &            NNPDF23LO            \\ \midrule
	$tttW$, $\ttbar\ttbar$ &                             \MGMCatNLO 2.2.2                             &    Full     &                                                       NLO~\cite{Alwall:2014hca}                                                       &               A14                &            NNPDF23LO            \\
	                             &                             + \PYTHIA 8.186                              &                &                                                                                                                                       &                                  &  \\ \midrule
	\ttbar                       &            \POWHEG v2 + \PYTHIA 6.428~\cite{Sjostrand:2006za}            &    Full     &                                                NNLO+NNLL~\cite{ATL-PHYS-PUB-2016-004}                                                 & Perugia2012~\cite{Skands:2010ak} &              CT10               \\ \midrule
	\Zjets, \Wjets               &                             \MGMCatNLO 2.2.2                             &    Full     &                                                   NNLO~\cite{ATL-PHYS-PUB-2016-003}                                                   &               A14                &            NNPDF23LO            \\
	                             &                             + \PYTHIA 8.186                              &                &                                                                                                                                       &                                  &  \\ \midrule
	SUSY signal                  &                             \MGMCatNLO 2.2.2                             &     Fast      & NLO+NLL~\cite{Beenakker:1996ch,Kulesza:2008jb,Kulesza:2009kq,Beenakker:2009ha,Beenakker:2011fu,Fuks:2012qx,Fuks:2013vua,Fuks:2013lya} &               A14                &            NNPDF23LO            \\
	                             &                             + \PYTHIA 8.186                              &                &                                                                                                                                       &                                  &  \\ \bottomrule
\end{tabular} 
\caption{Summary of the simulated SM background samples used in this analysis, where $V=W,Z$, and includes off-shell contributions. ``Tune'' refers to the set of tuned parameters used by the generator. The sample marked with a {$\mathbf{\dagger}$} is used for a cross-check of yields and for studies of systematic uncertainties.
\label{tab:SMbkgMC}}
\end{table}

\section{Event selection}\label{sec:eventreco}
After the application of beam, detector and data-quality requirements, the total integrated luminosity considered in this analysis corresponds to $36.1\pm1.2~\ifb$.
Events recorded during stable data-taking conditions are used in the analysis if the reconstructed primary vertex has at least two tracks with transverse momentum $\pt>400~\MeV$ associated with it.
The primary vertex of an event is identified as the vertex with the highest $\Sigma \pt^2$ of associated tracks.

Preselected electrons are required to have $|\eta|<2.47$ and $\pt>7\GeV$, where the $\pt$ and $\eta$ are determined from the calibrated clustered energy deposits in the electromagnetic calorimeter and the matched ID track, respectively.
Electrons must satisfy ``loose'' criteria of the likelihood-based identification algorithm~\cite{ATLAS-CONF-2016-024},
with additional track requirements based on the innermost pixel layer. 
Preselected muons are reconstructed by combining tracks in the ID with tracks in the MS~\cite{Aad:2016jkr}, and are required to have $|\eta|<2.7$ and $\pt>5\GeV$. 
Muons must satisfy ``medium'' identification requirements based on the number of hits in the different ID and MS subsystems, and the significance of the charge-to-momentum ratio, defined in Ref.~\cite{Aad:2016jkr}. 
Events containing one or more muons that have a transverse impact parameter relative to the primary vertex $|d_0|>0.2\,$mm or a longitudinal impact parameter relative to the primary vertex $|z_0|>1\,$mm are rejected to suppress the cosmic-ray muon background. 

Jets are reconstructed with the anti-$k_t$ algorithm~\cite{Cacciari:2008gp} with a radius parameter of $R=0.4$. 
Three-dimensional calorimeter energy clusters are used as input to the jet reconstruction, and jets are calibrated following Ref.~\cite{ATL-PHYS-PUB-2015-015}. 
Jets must have $|\eta|<2.8$ and $\pt>20\GeV$. 
To reduce pileup effects, jets with $\pt<60\GeV$ and $|\eta|<2.4$ must satisfy additional criteria using the jet vertex tagging algorithm described in Ref.~\cite{PERF-2014-03}.
Events containing jets failing to satisfy the quality criteria described in Ref.~\cite{Run2JetCalib} are rejected to suppress events with large calorimeter noise or non-collision backgrounds. 

The visible part of hadronically decaying tau leptons, denoted as $\tauhad$ and conventionally referred to as $\emph{taus}$ throughout this paper, is reconstructed~\cite{ATLAS-CONF-2017-029} using jets as described above with $|\eta|<2.47$ and $\pt>10\GeV$. 
The $\tauhad$ reconstruction algorithm uses information about the tracks within $\Delta R \equiv \sqrt{(\Delta\phi)^2+(\Delta\eta)^2}=0.2$ of the jet direction, in addition to the electromagnetic and hadronic shower shapes in the calorimeters. 
Preselected $\tauhad$ candidates are required to have one or three associated tracks (prongs), because taus predominantly decay to either one or three charged hadrons together with a neutrino and often additional neutral hadrons. 
The preselected $\tauhad$ are required to have $\pt>20\GeV$ and unit total charge of their constituent tracks.
In order to suppress electrons misidentified as preselected $\tauhad$, taus are vetoed using transition radiation and calorimeter information. 
The preselected $\tauhad$ candidates are corrected to the $\tauhad$ energy scale using an $\eta$- and $\pt$-dependent calibration.
A boosted decision tree algorithm (BDT) uses discriminating track and cluster variables to optimize $\tauhad$ identification,
where ``loose'', ``medium'' and ``tight'' working points are defined~\cite{ATL-PHYS-PUB-2015-045}, but not used to preselect tau leptons.
In this analysis, kinematic variables built with hadronically decaying taus use only their visible decay products.

The missing transverse momentum, $\met$, is the magnitude of the negative vector sum of the transverse momenta of all identified physics objects (electrons, photons, muons and jets) and an additional soft term~\cite{Run2MET}.
Taus are included as jets in the $\met$. 
The soft term is constructed from the tracks matched to the primary vertex, but not associated with identified physics objects, which allows the soft term to be nearly independent of pileup.

To avoid potential ambiguities among identified physics objects, preselected charged leptons and jets must survive ``overlap removal'', applied in the following order:
\vspace{-5pt}
\begin{enumerate}[nolistsep]
\item Any tau within $\Delta R=0.2$ of an electron or muon is removed.
\item Any electron sharing an ID track with a muon is removed.
\item Jets within $\Delta R=0.2$ of a preselected electron are discarded.
\item Electrons within $\Delta R=0.4$ of a preselected jet are discarded, to suppress electrons from semileptonic decays of $c$- and $b$-hadrons. 
\item Jets with fewer than three associated tracks are discarded either if a preselected muon is within $\Delta R=0.2$ or if the muon can be matched to a track associated with the jet. 
\item Muons within $\Delta R=0.4$ of a preselected jet are discarded to suppress muons from semileptonic decays of $c$- and $b$-hadrons. 
\item Jets within $\Delta R=0.4$ of a preselected tau passing ``medium'' identification requirements  are discarded.
\end{enumerate}
 Finally, to suppress low-mass particle decays, if surviving electrons and muons form an opposite-sign (OS) pair with $\mos < 4\GeV$, or form a same-flavor, 
 opposite-sign (SFOS) pair in the $\Upsilon(1S)-\Upsilon(3S)$ mass range $8.4<\msfos<10.4\GeV$, both leptons are discarded.

``Signal'' light charged leptons, abbreviated as \textit{signal leptons}, are preselected leptons surviving overlap removal and satisfying additional identification criteria. 
Signal electrons and muons must pass $\pt$-dependent isolation requirements, to reduce the contributions from semileptonic decays of hadrons and jets misidentified as prompt leptons. 
The isolation requirements use calorimeter- and track-based information to obtain 95\% efficiency for charged leptons with $\pT=25\GeV$ in $Z\ra\ee,\mumu$ events,
rising to 99\% efficiency at $\pt=60\GeV$.
To improve the identification of closely spaced charged leptons (e.g.\ from boosted decays), contributions to the isolation from nearby electrons and muons passing all other signal lepton requirements are removed. 
To further suppress electrons and muons originating from secondary vertices, $|z_{0}\sin{\theta}|$ is required to be less than $0.5\,$mm, and the $d_0$ normalized to its uncertainty is required to be small, with $|d_{0}|/\sigma_{d_{0}}<5 (3)$ for electrons (muons). 
Signal electrons must also satisfy ``medium'' likelihood-based identification criteria~\cite{ATLAS-CONF-2016-024}, 
while signal taus must satisfy the ``medium'' BDT-based identification criteria against jets~\cite{ATL-PHYS-PUB-2015-045}.

Events are selected using single-lepton or dilepton triggers, 
where the trigger efficiencies are in the plateau region above the offline $\pt$ thresholds indicated in Table~\ref{tab:triggers}. 
Dilepton triggers are used only when the leptons in the event fail $\pt$-threshold requirements for the single-lepton triggers.
The triggering efficiency for events with four, three and two electrons/muons in signal SUSY scenarios is typically $>$99\%, 96\% and 90\%, respectively.

\begin{table}[h]
  \begin{center}
  \small{
    \begin{tabular}{ l ll }
    	\toprule
    	Trigger                            & \multicolumn{2}{l}{Offline $\pt$ threshold [$\GeV$] } \\
    	                                   & 2015              & 2016                              \\ \midrule
    	Single isolated $e$                & 25                & 27                                \\
    	Single non-isolated $e$            & 61                & 61                                \\
    	Single isolated $\mu$              & 21                & 25 or 27                          \\
    	Single non-isolated $\mu$          & 41                & 41 or 51                          \\ \midrule
    	Double $e$                         & 13, 13            & 18, 18                            \\ \midrule
    	Double $\mu$ { \textit (symmetric)}     & --                & 11, 11 or 15, 15                  \\
    	~~~~~~~~~~~~~~~~{\textit(asymmetric)} & 19, 9             & 21, 9 or 23, 9                    \\ \midrule
    	Combined $e\mu$                    & 8($e$), 25($\mu$) & 8($e$), 25($\mu$)                 \\ \bottomrule
    \end{tabular} 
  }       
  \caption{The triggers used in the analysis of 2015 and 2016 data. 
  	The offline $\pt$ thresholds are required only for reconstructed charged leptons which match to the trigger signatures. 
  	Trigger thresholds for data recorded in 2016 are higher than in 2015 due to the increase in beam luminosity, and ``or'' denotes a move to a higher-threshold trigger during data-taking.
  \label{tab:triggers} }
  \end{center}
\end{table}

\section{Signal regions}\label{sec:signalregions}

Events with four or more signal leptons ($e$, $\mu$, $\tauhad$) are selected and are classified according to the number of light signal leptons ($L$ = $e$, $\mu$) and signal taus ($T$) required:
at least four light leptons and exactly zero taus $4L0T$,
exactly three light leptons and at least one tau $3L1T$,
or exactly two light leptons and at least two taus $2L2T$.

Events are further classified according to whether they are consistent with a leptonic $Z$ boson decay or not. 
The $Z$ requirement selects events where any SFOS $LL$ pair combination has an invariant mass close to the $Z$ boson mass, in the range $81.2$--$101.2\GeV$. 
A second $Z$ candidate may be identified if a second SFOS $LL$ pair is present and satisfies $61.2 < m(LL) < 101.2 \GeV$.
Widening the low-mass side of the $m(LL)$ window used for the selection of a second $Z$ candidate increases GGM signal acceptance. 
The $Z$ veto rejects events where any SFOS lepton pair combination has an invariant mass close to the $Z$ boson mass, in the range $81.2$--$101.2\GeV$. 
To suppress radiative $Z$ boson decays into four leptons  (where a photon radiated from a $Z\ra\ell\ell$ decay converts to a second SFOS lepton pair)
the $Z$ veto also considers combinations of any SFOS $LL$ pair with an additional lepton (SFOS+$L$), or with a second SFOS $LL$ pair (SFOS+SFOS), and rejects events where either the SFOS+$L$ or SFOS+SFOS invariant mass lies in the range $81.2$--$101.2\GeV$.

In order to separate the SM background from SUSY signal, 
the \met and the effective mass of the event, $\meff$, are both used.
The $\meff$ is defined as the scalar sum of the \met, 
the $\pt$ of signal leptons and the $\pt$ of all jets with $\pt>40\GeV$. 
The $\pt>40\GeV$ requirement for jets aims to suppress contributions from pileup and the underlying event.
A selection using the $\meff$ rather than the $\met$ is particularly effective for the RPV SUSY scenarios, which produce multiple high-energy leptons (and in some cases jets), but only low to moderate $\met$ from neutrinos in the final state.
The chosen $\meff$ thresholds are found to be close to optimal for the RPV scenarios with different NLSPs considered in this paper. 

Two signal regions (SR) are defined with $4L0T$ and a $Z$ veto: a general, model-independent signal region (SR0A) with $\meff>600\GeV$, and a tighter signal region (SR0B) with $\meff>1100\GeV$, optimized for the RPV $\LLE 12k$ scenarios.
Two further SRs are defined with $4L0T$, a first and second $Z$ requirement as described above, and different selections on $\met$: a loose signal region (SR0C) with $\met>50\GeV$, and a tighter signal region (SR0D) with $\met>100\GeV$, optimized for the low-mass and high-mass higgsino GGM scenarios, respectively.
Finally, two SRs are optimized for the tau-rich RPV $\LLE i33$ scenarios: one with $3L1T$ where the tau has $\pt>30\GeV$, a $Z$ veto and $\meff>700\GeV$ (SR1), and a second with $2L2T$ where the taus have $\pt>30\GeV$, a $Z$ veto and $\meff>650\GeV$ (SR2). 
The signal region definitions are summarized in Table~\ref{tab:SR_defs}.

\begin{table}[h]
\newcolumntype{R}{>{$}r<{$}}
\newcolumntype{L}{>{$}l<{$}}
\newcolumntype{M}{R@{${}>{}$}L}	
  \centering
  \small{
    \begin{tabular}{ lcccc M c }
    	\toprule
    	Region & $N(e,\mu)$ & $N(\tauhad)$ & $\pt\left({\tauhad}\right)$ &     $Z$ boson      & \multicolumn{2}{c}{Selection} & Target         \\ \midrule
    	SR0A   &  $\geq4$   &     $=0$     &         $>20\GeV$          &        veto        & $\meff$ & $600\GeV$         & General        \\
    	SR0B   &  $\geq4$   &     $=0$     &         $>20\GeV$          &        veto        & $\meff$ & $1100\GeV$         & RPV $\LLE12k$  \\ \midrule
    	SR0C   &  $\geq4$   &     $=0$     &         $>20\GeV$          & require 1st \& 2nd & $\met$  & $50\GeV$           & higgsino GGM   \\
    	SR0D   &  $\geq4$   &     $=0$     &         $>20\GeV$          & require 1st \& 2nd & $\met$  & $100\GeV$          & higgsino GGM   \\ \midrule
    	SR1    &    $=3$    &   $\geq1$    &         $>30\GeV$          &        veto        & $\meff$ & $700\GeV$          & RPV $\LLE i33$ \\
    	SR2    &    $=2$    &   $\geq2$    &         $>30\GeV$          &        veto        & $\meff$ & $650\GeV$          & RPV $\LLE i33$ \\ \bottomrule
    \end{tabular} 
  }       
  \caption{Signal region definitions. 
  The $\pt\left({\tauhad}\right)$ column denotes the $\pt$ threshold used for the tau selection or veto.
  SR0B and SR0D are subsets of SR0A and SR0C, respectively, while SR1 and SR2 are completely disjoint.
\label{tab:SR_defs}}
\end{table}

\section{Background determination}\label{sec:bgdet}

Several SM processes can result in signatures resembling SUSY signals with four reconstructed charged leptons, 
including both the ``real'' and ``fake'' lepton contributions. 
Here, a real charged lepton is defined to be a prompt and genuinely isolated lepton, 
while a fake charged lepton is defined to be a non-prompt or non-isolated lepton that could originate from semileptonic decays of $b$- and $c$-hadrons, or from in-flight decays of light mesons, 
or from misidentification of particles within light-flavor or gluon-initiated jets, or from photon conversions.
The SM processes are classified into two categories:
\begin{description}
\item[Irreducible background:] hard-scattering processes giving rise to events with four or more real leptons, \ZZ, \ttZ, \ttWW, \tWZ, $VVZ$ (\ZZZ, \WZZ, \WWZ), Higgs ($ggH$, $WH$, $ZH$, $\ttbar H$), \tttt, $\ttbar t W$.
\item[Reducible background:] processes leading to events with at least one fake lepton, \ttbar, \Zjets, \WZ, \WW, \WWW, \ttW, $\ttbar t$. Processes listed under irreducible that do not undergo a decay to four real leptons (e.g.\ $ZZ\ra q\bar{q}\ell\ell$) are also included in the reducible background.
\end{description}
Backgrounds with three or more fake leptons (e.g.\ \Wjets) are found to be very small for this analysis, and the systematic uncertainty on the reducible background is increased to cover any effect from them (discussed in Section~\ref{sec:systematics}).

In the signal regions, the irreducible background is dominated by \ttZ, $VVZ$ ($V=W,Z$), and \ZZ, while the reducible background is dominated by the two-fake-lepton backgrounds \ttbar and \Zjets.
The irreducible backgrounds are estimated from MC simulation, while the reducible backgrounds are derived from data with the fake-factor method.
Signal regions with $4L0T$ are dominated by irreducible background processes, whereas the reducible background processes dominate the $3L1T$ and $2L2T$ signal regions. 
The predictions for irreducible and reducible backgrounds are tested in validation regions (Section~\ref{sec:bgdet_valid}).
In the fake-factor method, the number of reducible background events in a given region is estimated from data using probabilities for a fake preselected lepton to pass or fail the signal lepton selection. 
The ratio $F=f/{\bar{ f}}$ for fake leptons is the ``fake factor'', where $f$ ($\bar f$) is the probability that a fake lepton is misidentified as a signal (``loose'') lepton.
The probabilities used in the fake-factor calculations are based on simulation and corrected to data where possible.
Loose leptons are preselected leptons surviving overlap removal that do not satisfy signal lepton criteria.
For this fake-factor evaluation, a very loose selection on the identification BDT is also applied to the preselected taus, 
since candidates with very low BDT scores are typically gluon-induced jets and jets arising from pileup, which is not the case for the signal tau candidates.

The reducible background prediction is extracted by applying fake factors to control regions (CR) in data. 
The CR definition only differs from that of the associated SR in the quality of the required leptons; here exactly one (CR1) or two (CR2) of the four leptons must be identified as a loose lepton, as shown in Table~\ref{tab:CR_defs}. 
In 3$L$1$T$ events, the contribution from events with two fake light leptons is negligible, as is the contribution from one and two fake light leptons in 2$L$2$T$ events.

\begin{table}[h]
  \centering
  \small{
    \begin{tabular}{ ll cccc }
    	\toprule
    	Reducible      & Control Region & $N(e,\mu)$ & $N(e,\mu)$ & $N(\tauhad)$ & $N(\tauhad)$ \\
    	Estimation for &                & signal     &   loose    &    signal    &    loose     \\ \midrule
    	$4L0T$         & CR1\_LLLl      & $=3$       &  $\geq1$   &     $=0$     &   $\geq0$    \\
    	               & CR2\_LLll      & $=2$       &  $\geq2$   &     $=0$     &   $\geq0$    \\ \midrule
    	$3L1T$         & CR1\_LLLt      & $=3$       &    $=0$    &     $=0$     &   $\geq1$    \\
    	               & CR1\_LLTl      & $=2$       &    $=1$    &   $\geq1$    &   $\geq0$    \\
    	               & CR2\_LLlt      & $=2$       &    $=1$    &     $=0$     &   $\geq1$    \\ \midrule
    	$2L2T$         & CR1\_LLTt      & $=2$       &    $=0$    &     $=1$     &   $\geq1$    \\
    	               & CR2\_LLtt      & $=2$       &    $=0$    &     $=0$     &   $\geq2$    \\ \bottomrule
    \end{tabular} 
  }       
  \caption{Control region definitions where ``L'' and ``T'' denote signal light leptons and taus, while ``l'' and ``t'' denote loose light leptons and taus.  
      Loose leptons are preselected leptons surviving overlap removal that do not pass signal lepton criteria.
      Additional selection for $\pt\left({\tauhad}\right)$,  $Z$ veto/requirement, $\met$, $\meff$ are applied to match a given signal or validation region.
\label{tab:CR_defs}}
\end{table}

Fake factors are calculated separately for fake electrons, muons and taus, from light-flavor jets, heavy-flavor jets, gluon-initiated jets (taus only) and photon conversions (electrons and taus only).
These categories are referred to as fake-lepton ``types''.
The fake factor for each fake-lepton type is computed for each background process due to a dependence on the hard process (e.g.\ \ttbar, \Zjets). 
The fake factor per fake-lepton type and per process is binned in lepton $\pt$, $\eta$ and number of prongs for taus. 

To account correctly for the relative abundances of fake-lepton types and production processes, a weighted average $F_{w}$ of fake factors is computed in each CR, as:

\vspace{-0.4cm}
\begin{equation*}
\label{eq:fr}
F_{w}=\sum_{i,j} \left(  R^{ij} \times s^{i} \times F^{ij} \right).
\end{equation*}

\vspace{-0.3cm}
\noindent The factors $R^{ij}$ are ``process fractions'' that depend on the fraction of fake leptons of type $i$ from process $j$, determined from MC simulation in the corresponding CR2, 
and are similar to the process fractions obtained in the signal regions from MC simulation, which suffer from having few events. 
The term $F^{ij}$ is the corresponding fake factor calculated using MC simulation. 
The ``scale factors'' $s^{i}$ are corrections that depend on the fake-lepton type, and are applied to the fake factors to account for possible differences between data and MC simulation. 
These are assumed to be independent of the physical process, 
and are determined from data in dedicated regions enriched in objects of a given fake-lepton type. 

For fake light leptons from heavy-flavor jets, the scale factor is measured in a $\ttbar$-dominated control sample. 
The heavy-flavor scale factors are seen to have a modest $\pt$-dependence, 
decreasing for muons from $ 1.00\pm 0.07$ to $0.73\pm 0.18$ as the muon $\pt$ increases from $5\GeV$ to $20\GeV$.
For electrons, the heavy-flavor scale factor is seen to increase from $ 1.16\pm 0.11$ to $1.35\pm 0.29$ across the same $\pt$ range. 
For taus, the heavy-flavor, gluon-initiated and conversion scale factors cannot be reliably measured using data.
Instead, they are assumed to be the same as the light-flavor jet scale factor described below. 

The scale factor for fake taus originating from light-flavor jets is measured separately for one- and three-prong taus in a control sample dominated by $Z$+jets events. 
The scale factors are seen to be $\pt$-dependent, 
decreasing from $ 1.30\pm 0.05$ to $0.96\pm 0.06$ ($ 1.42\pm 0.11$ to $1.23\pm 0.13$) as the 1-prong (3-prong) tau $\pt$ increases from $20\GeV$ to $60\GeV$. 
The contribution to the signal regions from fake light leptons originating from light-flavour jets is very small (less than $1.8\%$ of all $e,\mu$) and the scale factor cannot be reliably measured using data.
Therefore, values of $1.00 \pm 0.25$ are used instead, motivated by similar uncertainties in the other scale factor measurements.

For fake electrons from conversions, the scale factor is determined in a sample of photons from final-state radiation of $Z$ boson decays to muon pairs.
The electron conversion scale factor is seen to have a small $\pt$-dependence, increasing from $ 1.38 \pm 0.17$ to $1.53 \pm 0.20$ as the electron $\pt$ increases from $7$ to $25\GeV$.

The number $N^\textrm{SR}_\textrm{red}$ of background events with one or two fake leptons from reducible sources in each SR is determined from the number of events in data in the corresponding CRs, $N^\textrm{CR1}_\textrm{data}$ and $N^\textrm{CR2}_\textrm{data}$, according to:

\vspace{-0.4cm}
\begin{equation}
\label{eq:fakes}
N^\textrm{SR}_\textrm{red}  = [N^\textrm{CR1}_\textrm{data}-N^\textrm{CR1}_\textrm{irr}]\times F_{w,1} - [N^\textrm{CR2}_\textrm{data}-N^\textrm{CR2}_\textrm{irr}]\times F_{w,1}\times F_{w,2},
\end{equation}

\vspace{-0.2cm}
\noindent where $F_{w,1}$ and $F_{w,2}$ are the two weighted fake factors that are constructed using the leading and subleading in $\pt$ loose leptons in the CRs, respectively.
The small contributions from irreducible background processes in the CRs, $N^\textrm{CR1,CR2}_\textrm{irr}$, are evaluated using MC simulation and subtracted from the corresponding number of events seen in data.
The second term removes the double-counting of events with two fake leptons in the first term.
Both CR1 and CR2 are dominated by the two-fake-lepton processes $\ttbar$ and $Z$+jets, thus the first term is roughly double the second term.
Higher-order terms in $F_w$ describing three- and four-fake-lepton backgrounds are neglected, as are some terms with a very small contribution; e.g.\ in 3$L$1$T$ events, the contribution from events with two fake light leptons is negligible.
A systematic uncertainty is applied to account for these neglected terms, as described in the following section.

\subsection{Systematic uncertainties \label{sec:systematics} }

Several sources of systematic uncertainty are considered for the SM background estimates and signal yield predictions. 
The systematic uncertainties affecting the simulation-based estimate can be divided into three components: MC statistical uncertainty, sources of experimental uncertainty (from identified physics objects $e$, $\mu$, $\tau$ and jets, and also $\met$), and sources of theoretical uncertainty.
The reducible background is affected by different sources of uncertainty associated with data counts in control regions and uncertainties in the weighted fake factors. 
The primary sources of systematic uncertainty, described below, are summarized in Figure~\ref{fig:systUncert}.

\begin{figure}[h]
\centering
\includegraphics[width=0.75\textwidth]{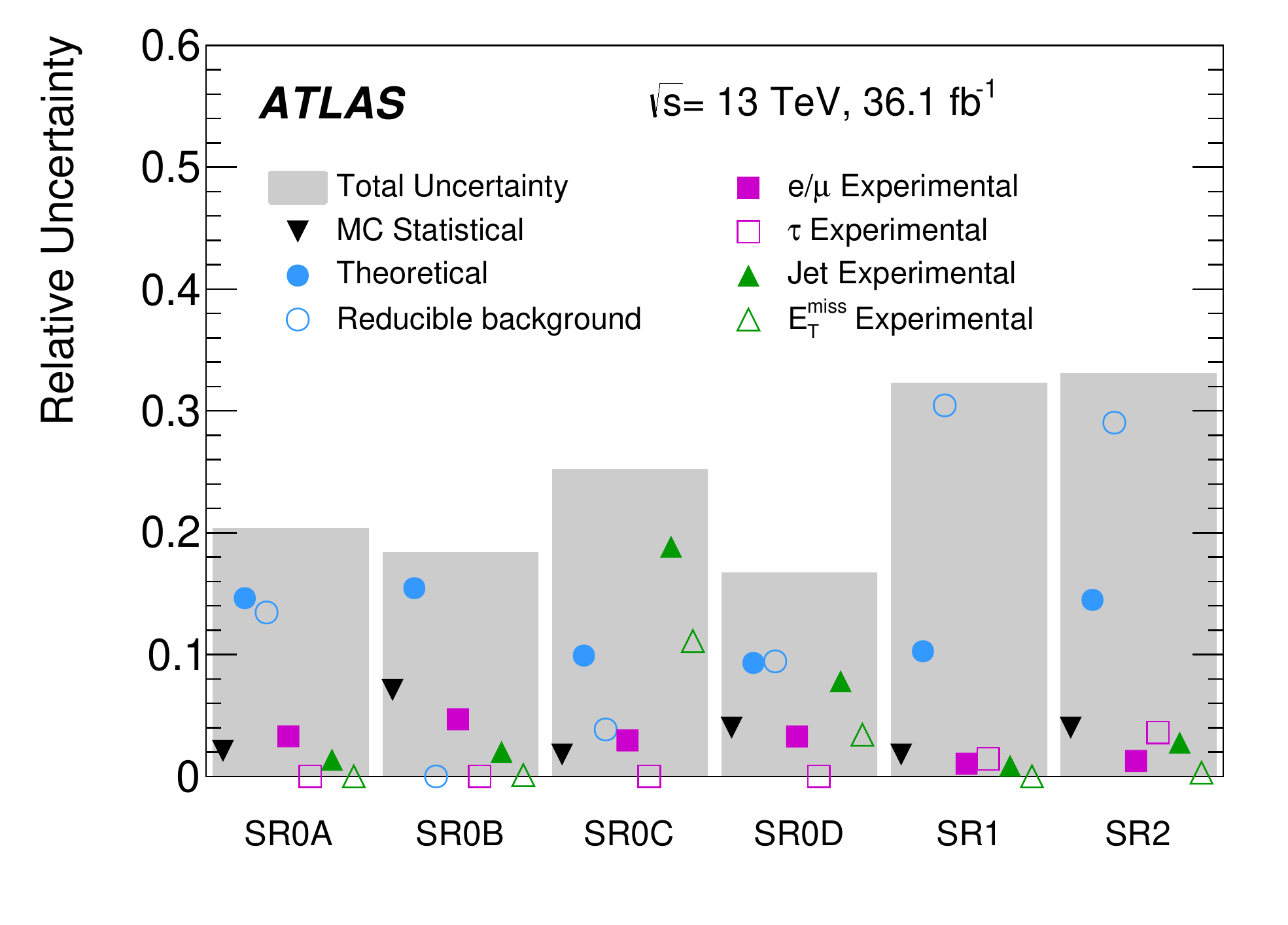}
\caption{Breakdown of the dominant systematic uncertainties in the background estimates for the signal regions. The individual uncertainties can be correlated, and do not necessarily sum in quadrature to the total background uncertainty. The text provides category details. 
\label{fig:systUncert}
}
\end{figure}

The MC statistical uncertainty for the simulation-based background estimate is small and less than $7\%$ of the total background estimate in all signal regions. 
Systematic uncertainties in the SUSY signal yields from experimental and theoretical sources are typically of the order of 10\% each. 
The experimental uncertainties include
the uncertainties associated with electrons, muons, taus, jets, and $\met$, 
as well as the uncertainty associated with the simulation of pileup,
and uncertainty in the luminosity ($2.1$\%, following a methodology similar to that detailed in Ref.~\cite{DAPR-2011-01}).
The uncertainties associated with pileup and luminosity are included in the total uncertainty in Figure~\ref{fig:systUncert}. 
The experimental uncertainties pertaining to electrons, muons and taus include the uncertainties due to the 
lepton identification efficiencies, lepton energy scale and energy resolution, isolation and trigger efficiencies.
Systematic uncertainties from electron, muon, and tau sources are generally low in all signal regions, at about 5\% relative to the total expected background. 
The uncertainties associated with jets are due to the jet energy scale, jet energy resolution and jet vertex tagging.
Uncertainties in the object momenta are propagated to the \met measurement, and additional uncertainties in $\met$
arising from energy deposits not associated with any reconstructed objects are also considered. 
The jet and $\met$ uncertainties are generally of the order of a few percent in the signal regions, 
but this rises to $21\%$ ($7\%$) in SR0C (SR0D), where a selection on $\met$ is made.

Theoretical uncertainties in the simulation-based estimates include   
the theoretical cross section uncertainties due to the choice of renormalization and factorization scales and PDFs,  
the acceptance uncertainty due to PDF and scale variations, and the choice of MC generator. 
The theoretical cross section uncertainties for the irreducible backgrounds used in this analysis are 
12\% for \ttZ~\cite{ATL-PHYS-PUB-2016-005}, 
6\% for \ZZ~\cite{ATL-PHYS-PUB-2016-002}, 
and 20\% for the triboson samples~\cite{ATL-PHYS-PUB-2016-002}, where the order of the cross section calculations is shown in Table~\ref{tab:SMbkgMC}. 
For the Higgs boson samples, an uncertainty of 20\% is used for $WH$, $ZH$ and VBF~\cite{deFlorian:2016spz}, 
while uncertainties of 100\% are assigned to $\ttbar H$ and $ggH$~\cite{Dittmaier:2012vm}. 
The uncertainties in the $\ttbar H$ and $ggH$ estimates are assumed to be large to account for uncertainties in the acceptance, while the inclusive cross sections are known to better precision. 
Uncertainties arising from the choice of generator are determined by comparing the \MGMCatNLO and \SHERPA generators for \ttZ. 
Finally, the uncertainty in the \ZZ and \ttZ acceptance due to PDF variations, and due to varying the renormalization and factorization scales by factors of $1/2$ and $2$, is also taken into account.
In SR0A and SR0B, the theoretical uncertainties dominate the total uncertainty, mainly due 
to the $20\%$ uncertainty from the $\ttZ$ MC generator choice, and the $10\%$ uncertainty from the $\ttZ$ PDF/scale variations ($25\%$ for $ZZ$).

The uncertainty in the reducible background is dominated by the statistical uncertainty of the data events in the corresponding CR1 and CR2. 
The uncertainty in the weighted fake factors includes the MC statistical uncertainty in the process fractions, the uncertainty in the fake lepton scale factors, and the statistical uncertainty from the fake factors measured in simulation.
The uncertainties for the fake factors from each fake-lepton type are treated as correlated across processes. 
Thus, since both CR1 and CR2 are dominated by two-fake-lepton processes with the same type of fake lepton, correlations in the fake factors applied to CR1 and CR2 result in a close cancelation of the uncertainties from the weighted fake factors between the first and second terms in Eq.~\eqref{eq:fakes}. 
Finally, a conservative uncertainty is applied to account for the neglected terms in Eq.~\eqref{eq:fakes}.
For example, in 4$L$0$T$ events the three- and four-fake-lepton terms are neglected.
Weighted fake factors are applied to data events with one signal and three loose light leptons to estimate an upper limit on this neglected contribution for each 4$L$0$T$ validation region (VR) and SR.
The calculated upper limit plus $1\sigma$ statistical uncertainty is added to the reducible background uncertainty, adding an absolute uncertainty of $0.14$ events in SR0A. 
This is repeated for the 3$L$1$T$ and 2$L$2$T$ regions, accounting for the neglected terms with one or two fake light leptons as necessary, adding an absolute uncertainty of $0.07$ events in SR1, and $0.20$ events in SR2.

\subsection{Background modeling validation}\label{sec:bgdet_valid}

The general modeling of both the irreducible and reducible backgrounds is tested in VRs that are defined to be adjacent to, yet disjoint from, the signal regions, as shown in Table~\ref{tab:VR_defs}.
For signal regions that veto $Z$ boson candidates, three VRs are defined by reversing the $\meff$ requirement, while for signal regions requiring two $Z$ boson candidates, one VR is defined by vetoing the presence of a second $Z$ boson candidate. 
The background model adopted in the VRs is the same as in the SRs, with the irreducible backgrounds obtained from MC simulation and the reducible background estimated from data using the fake-factor method with process fractions and loose lepton control regions corresponding to the VRs.
The systematic uncertainties on the SM backgrounds in the VRs are evaluated as in Section~\ref{sec:systematics}. 
The SM background in the VRs is dominated by \ZZ, \ttbar and \Zjets.

\begin{table}[h]
  \centering
  \small{
    \begin{tabular}{ l ccc c cc }
    	\toprule
    	Validation & $N(e,\mu)$ & $N(\tauhad)$ & $\pt\left({\tauhad}\right)$ & $Z$ boson               & Selection        &         Target         \\ 
    	Region     &            &              &                             &                         &                  &  \\ \midrule
    	VR0        & $\geq4$    &     $=0$     &         $>20\GeV$          & veto                    & $\meff<600\GeV$ & $\ttbar$, \Zjets, $ZZ$ \\
    	VR0Z       & $\geq4$    &     $=0$     &         $>20\GeV$          & require 1st \& veto 2nd & --               &          $ZZ$          \\
    	VR1        & $=3$       &   $\geq1$    &         $>30\GeV$          & veto                    & $\meff<700\GeV$ &    $\ttbar$, \Zjets    \\
    	VR2        & $=2$       &   $\geq2$    &         $>30\GeV$          & veto                    & $\meff<650\GeV$ &    $\ttbar$, \Zjets    \\ \bottomrule
    \end{tabular} 
  }       
  \caption{Validation region definitions.
    The $\pt\left({\tauhad}\right)$ column denotes the $\pt$ threshold used for the tau selection or veto.
    \label{tab:VR_defs}}
\end{table}

Observed and expected event yields in the VRs are shown in Table~\ref{tab:VRyields}, where good agreement is seen in general within statistical and systematic uncertainties. 
No significant excesses above the SM expectations are observed in any VR.

  \begin{table}[bh]
  	\newcolumntype{R}{>{$}r<{$}}
  	\newcolumntype{L}{>{$}l<{$}}
  	\newcolumntype{M}{R@{${}\pm{}$}L}
  	\centering
  	\small{
  		\begin{tabular}{ l M M M M } 
  			\toprule
  			Sample       &  \multicolumn{2}{c}{VR0}  & \multicolumn{2}{c}{ VR0Z}  &  \multicolumn{2}{c}{VR1}  & \multicolumn{2}{c}{VR2}  \\ \midrule
  			Observed     & \multicolumn{2}{c}{$132$} & \multicolumn{2}{c}{ $365$} & \multicolumn{2}{c}{$116$} & \multicolumn{2}{c}{$32$} \\ \midrule
  			SM Total     & $123 $    & $ 11$         & $334 $    & $ 52$          & $91 $      & $ 19$        & $28 $      & $ 6$        \\ \midrule
  			$ZZ$         & $65 $     & $ 7$          & $234 $    & $ 23$          & $8.8 $     & $ 1.0$       & $3.4 $     & $ 0.5$      \\
  			$\ttbar Z$   & $3.9 $    & $ 0.6$        & $10.5 $   & $ 1.5$         & $1.76 $    & $ 0.25$      & $0.60 $    & $ 0.10$     \\
  			Higgs        & $5 $      & $ 4$          & $43 $     & $ 37$          & $3.2$      & $ 2.9$       & $1.3 $     & $ 1.2$      \\
  			$VVV$        & $2.9 $    & $ 0.6$        & $16.1 $   & $ 3.4$         & $1.23 $    & $ 0.27$      & $0.29 $    & $ 0.07$     \\
  			Reducible    & $46 $     & $ 7$          & $28 $     & $ 26$          & $76 $      & $ 19$        & $22 $      & $ 6$        \\
  			Other        & $0.40 $   & $ 0.07$       & $2.7 $    & $ 0.5$         & $0.34 $    & $ 0.06$      & $0.16 $    & $ 0.04$     \\ \bottomrule
  		\end{tabular} 
  	}   
  \caption{Expected and observed yields for $36.1\,\ifb$ in the validation regions. ``Other'' is the sum of the \tWZ, \ttWW, and \tttt backgrounds.
  Both the statistical and systematic uncertainties in the SM background are included in the uncertainties shown. 
\label{tab:VRyields}}
\end{table}

The lepton \pt, $m_{\mathrm{SFOS}}$ and $\met$ distributions in the VRs are shown in Figure~\ref{fig:VR0LepPtAll} and Figure~\ref{fig:VR12LepPtAll}.
Figure~\ref{fig:VR0LepPt} shows that VR0 has a slight downward trend in the ratio of the data to estimated SM background as the \pt of the leptons increases, 
which was found to be most noticeable in the \pt of the leading electron in the event. 
However, since the corresponding signal regions (SR0A and SR0B) require high \meff, 
the potential impact of a small mismodeling of one electron in the event was found to be insignificant.

The \meff\ distributions in VR0, VR1 and VR2 can be seen in the lower \meff bins in Figure~\ref{fig:SRmeffmet}. 

\begin{figure}[h]
\centering
\subfloat[][$\pt\left({e,\mu}\right)$ in VR0]{\includegraphics[width=0.475\textwidth]{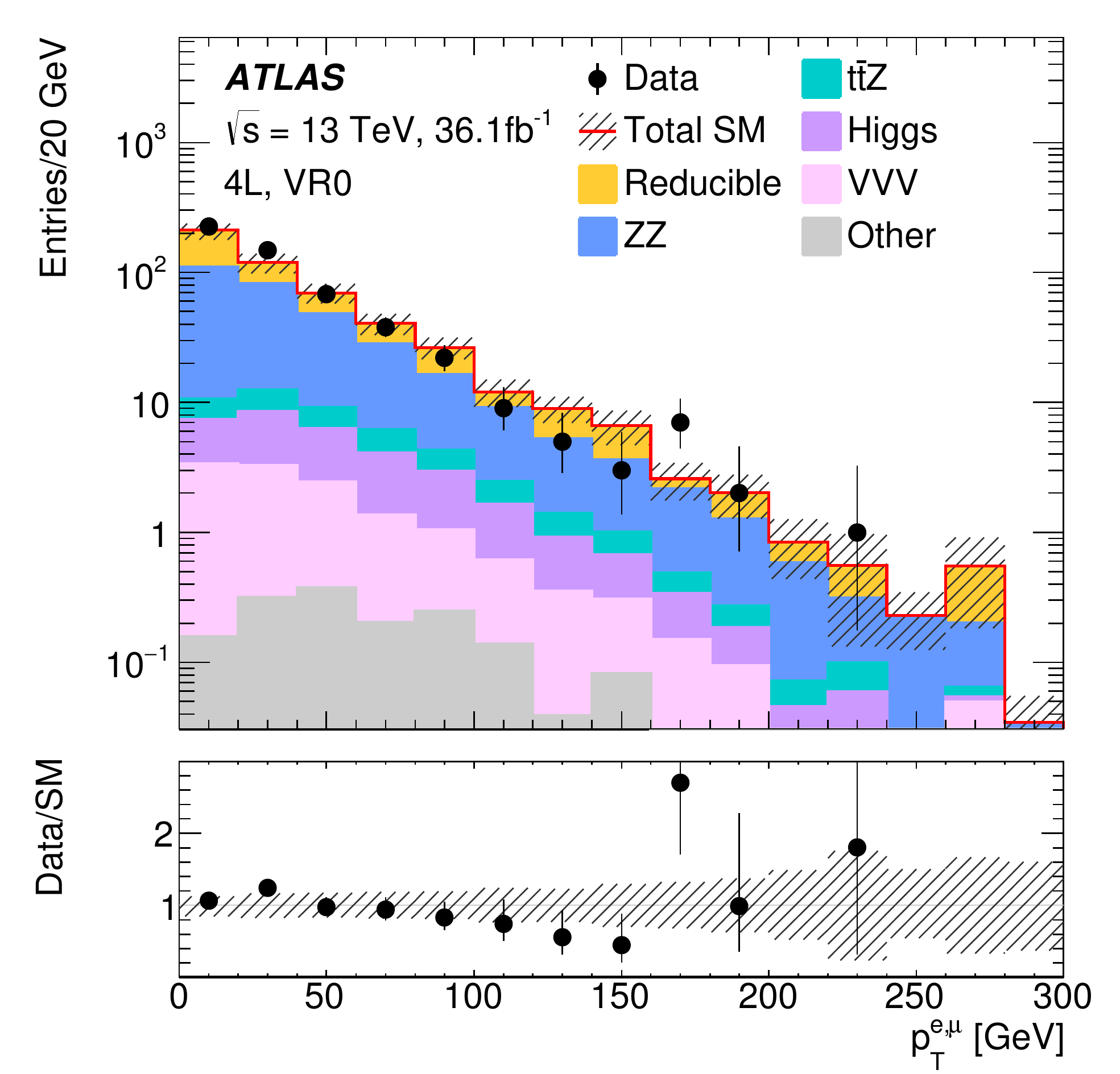}\label{fig:VR0LepPt}}
\subfloat[][$m_{\mathrm{SFOS}}$ in VR0]{\includegraphics[width=0.475\textwidth]{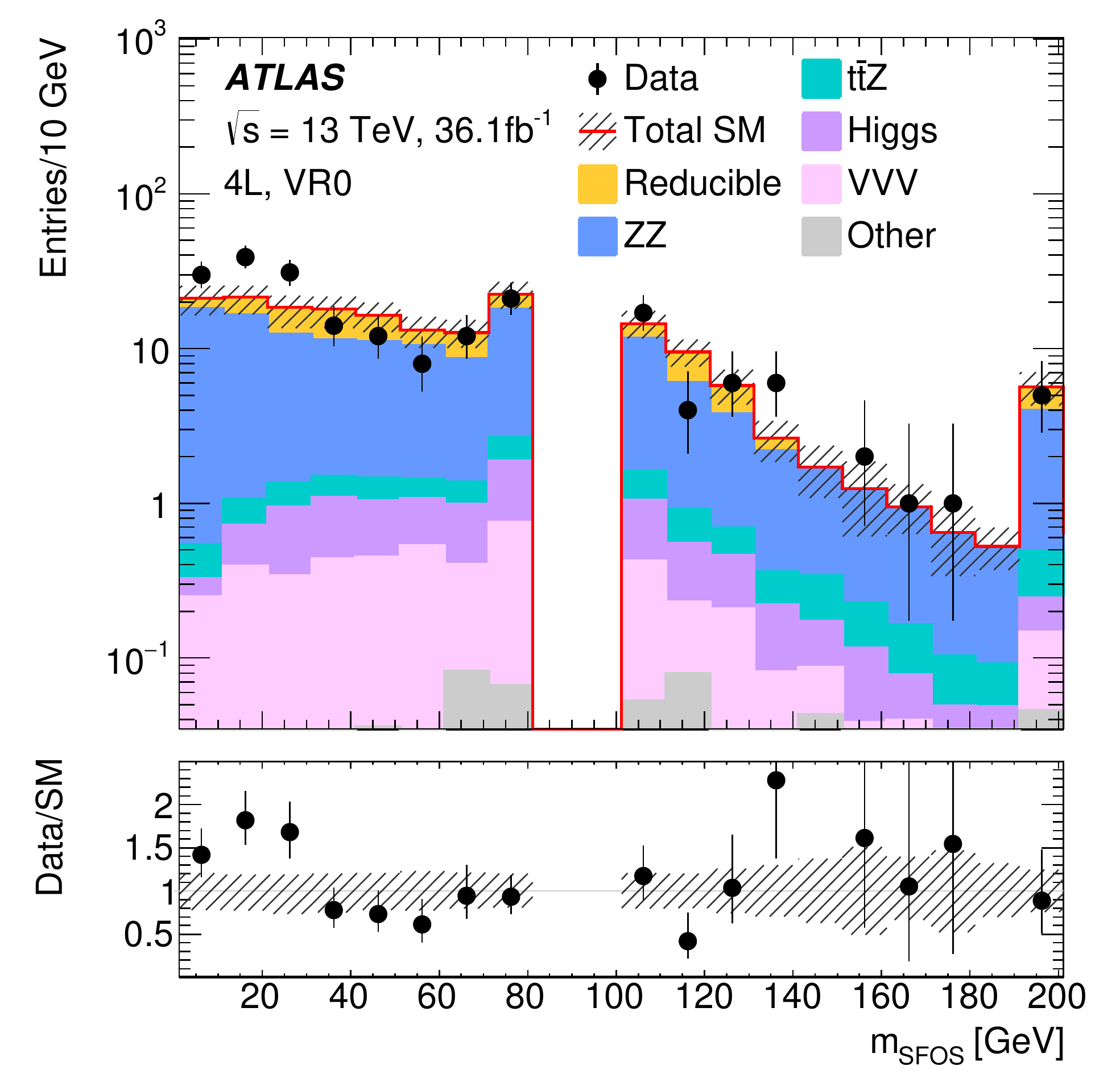}\label{fig:VR0mSFOS}}\\
\subfloat[][$\pt\left({e,\mu}\right)$ in VR0Z]{\includegraphics[width=0.475\textwidth]{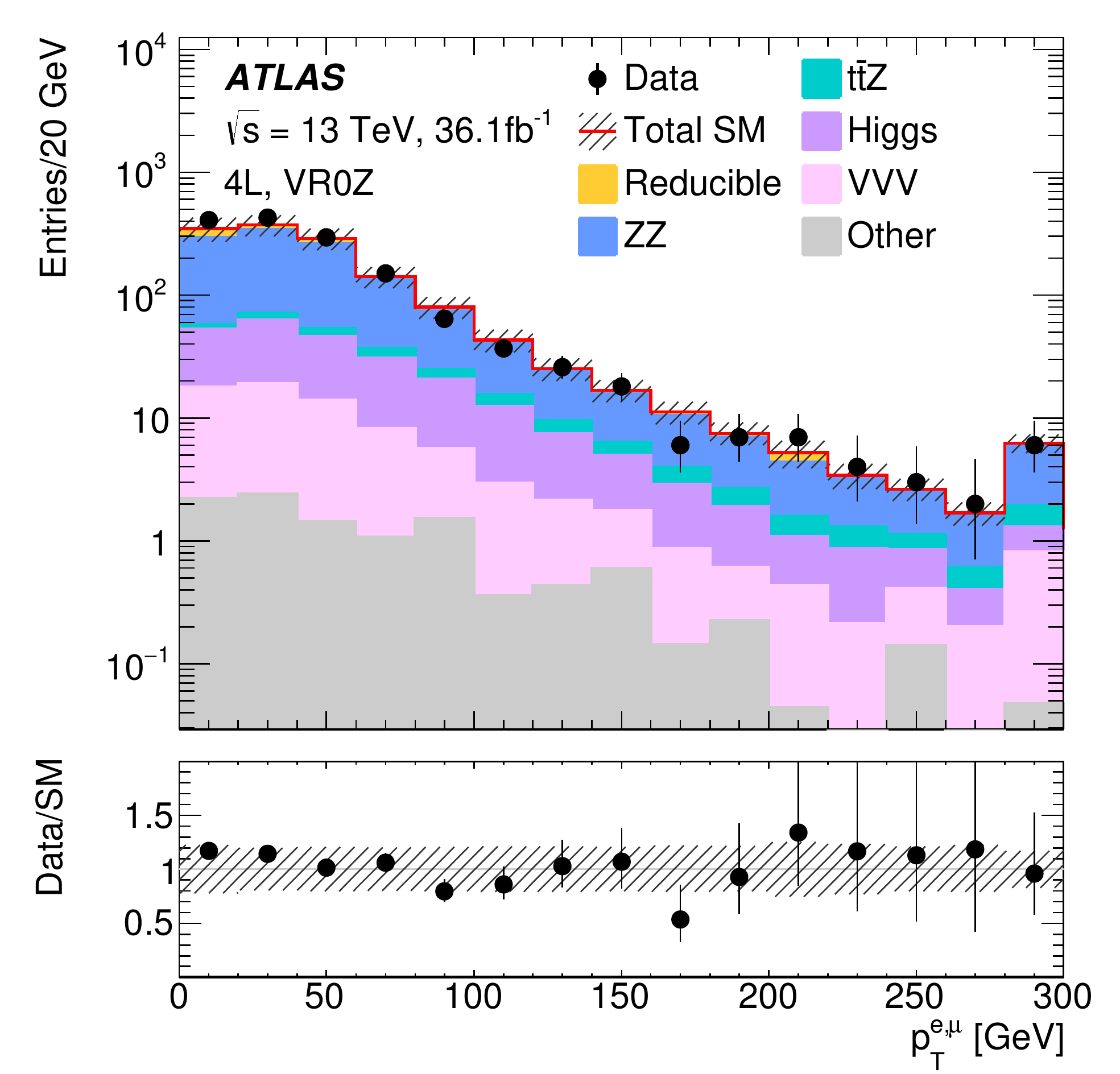}\label{fig:VR0ZLepPt}}
\subfloat[][$\met$ in VR0Z]{\includegraphics[width=0.475\textwidth]{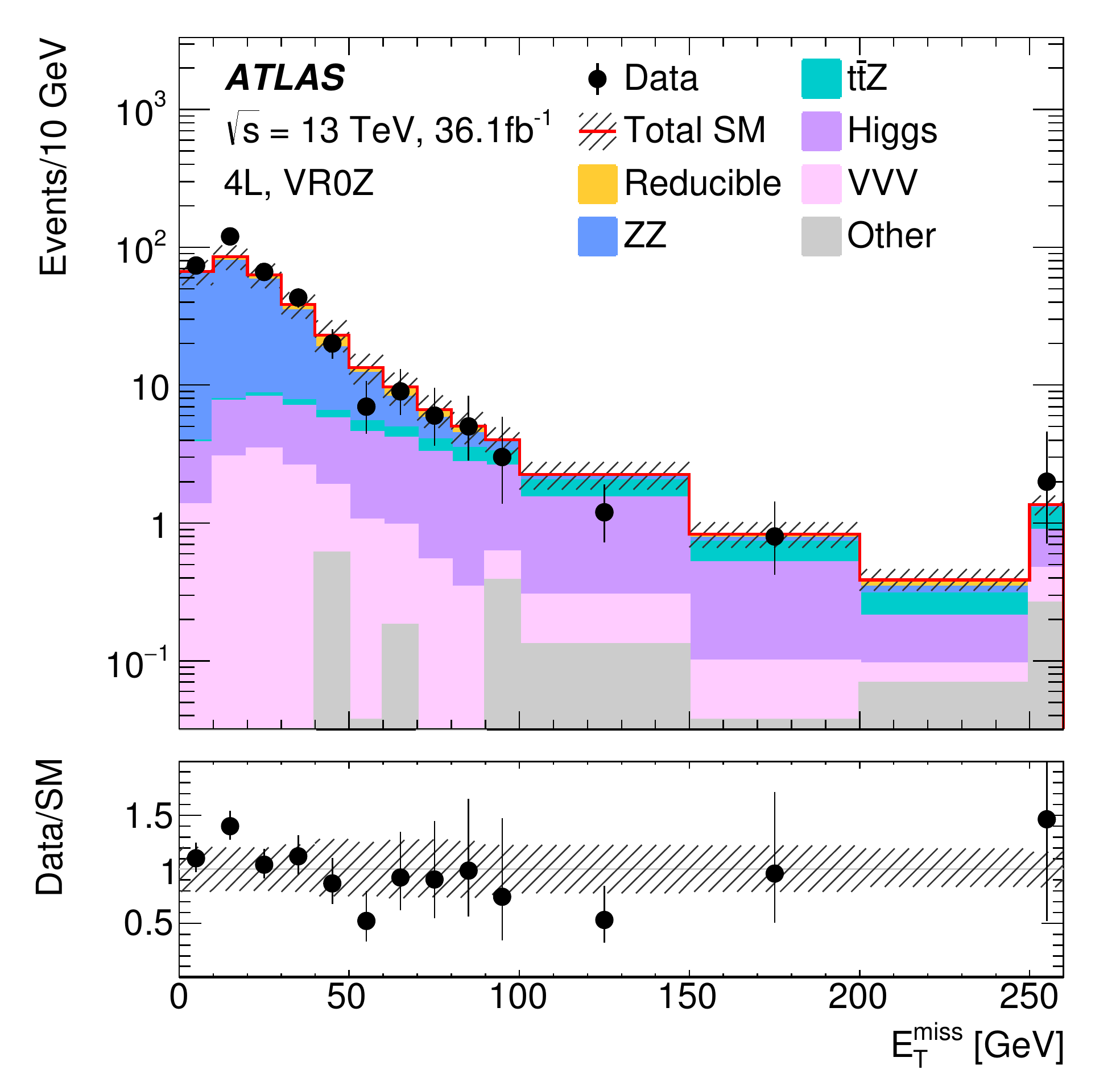}\label{fig:VR0ZMet}}
\caption{ The distributions for data and the estimated SM backgrounds in VR0 and VR0Z for 
\protect\subref{fig:VR0LepPt} \& \protect\subref{fig:VR0ZLepPt}~the electron and muon \pt, 
\protect\subref{fig:VR0mSFOS}~the SFOS invariant mass, and 
\protect\subref{fig:VR0ZMet}~the \met.
``Other'' is the sum of the \tWZ, \ttWW, and \tttt backgrounds.
The last bin includes the overflow. 
 Both the statistical and systematic uncertainties in the SM background are included in the shaded band.
\label{fig:VR0LepPtAll}
}
\end{figure}

\begin{figure}[h]
\centering
\subfloat[][$\pt\left({e,\mu}\right)$ in VR1]{\includegraphics[width=0.475\textwidth]{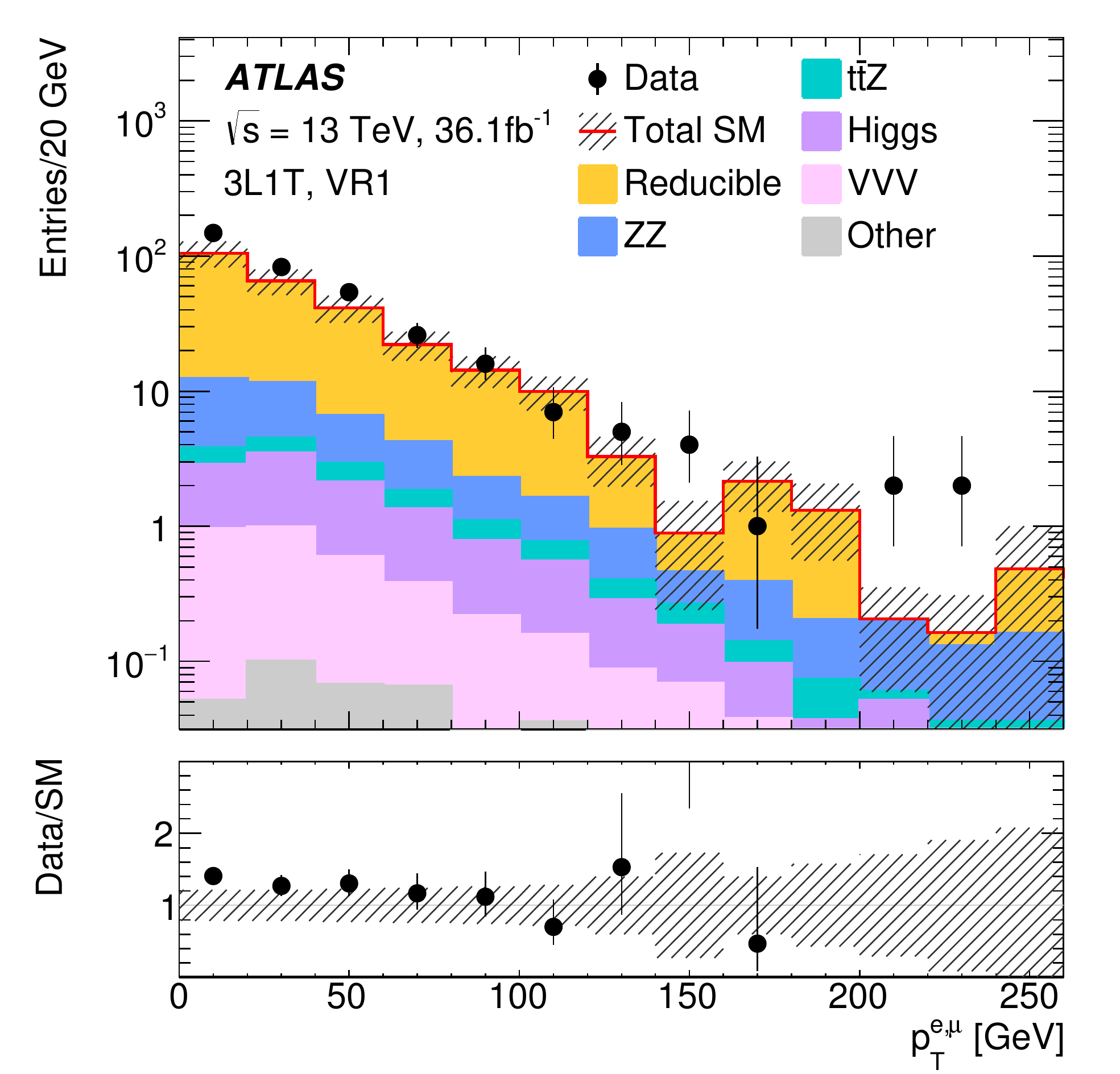}\label{fig:VR1LepPt}}
\subfloat[][$\pt\left({\tauhad}\right)$ in VR1]{\includegraphics[width=0.475\textwidth]{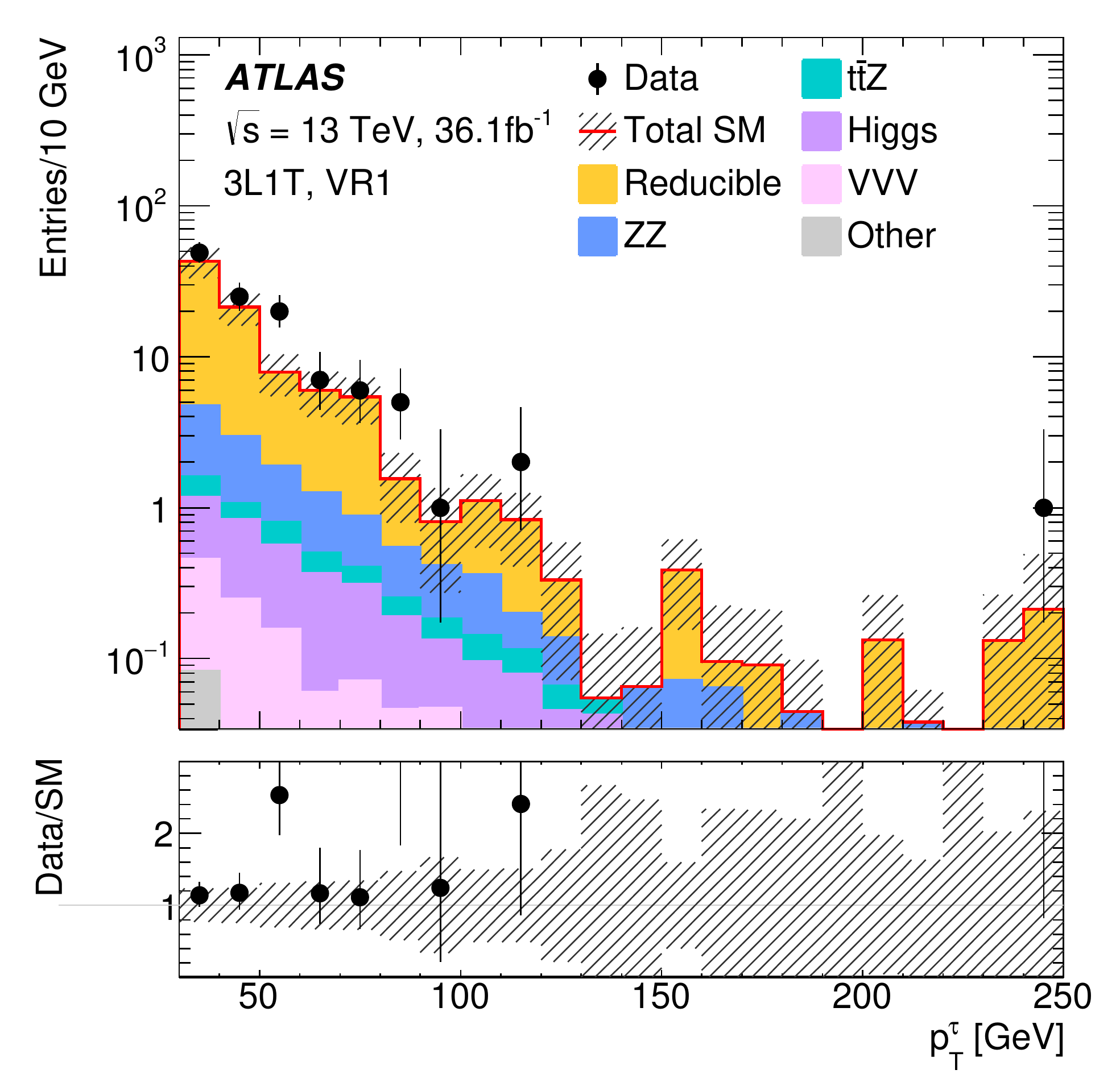}\label{fig:VR1TauPt}}\\
\subfloat[][$\pt\left({e,\mu}\right)$ in VR2]{\includegraphics[width=0.475\textwidth]{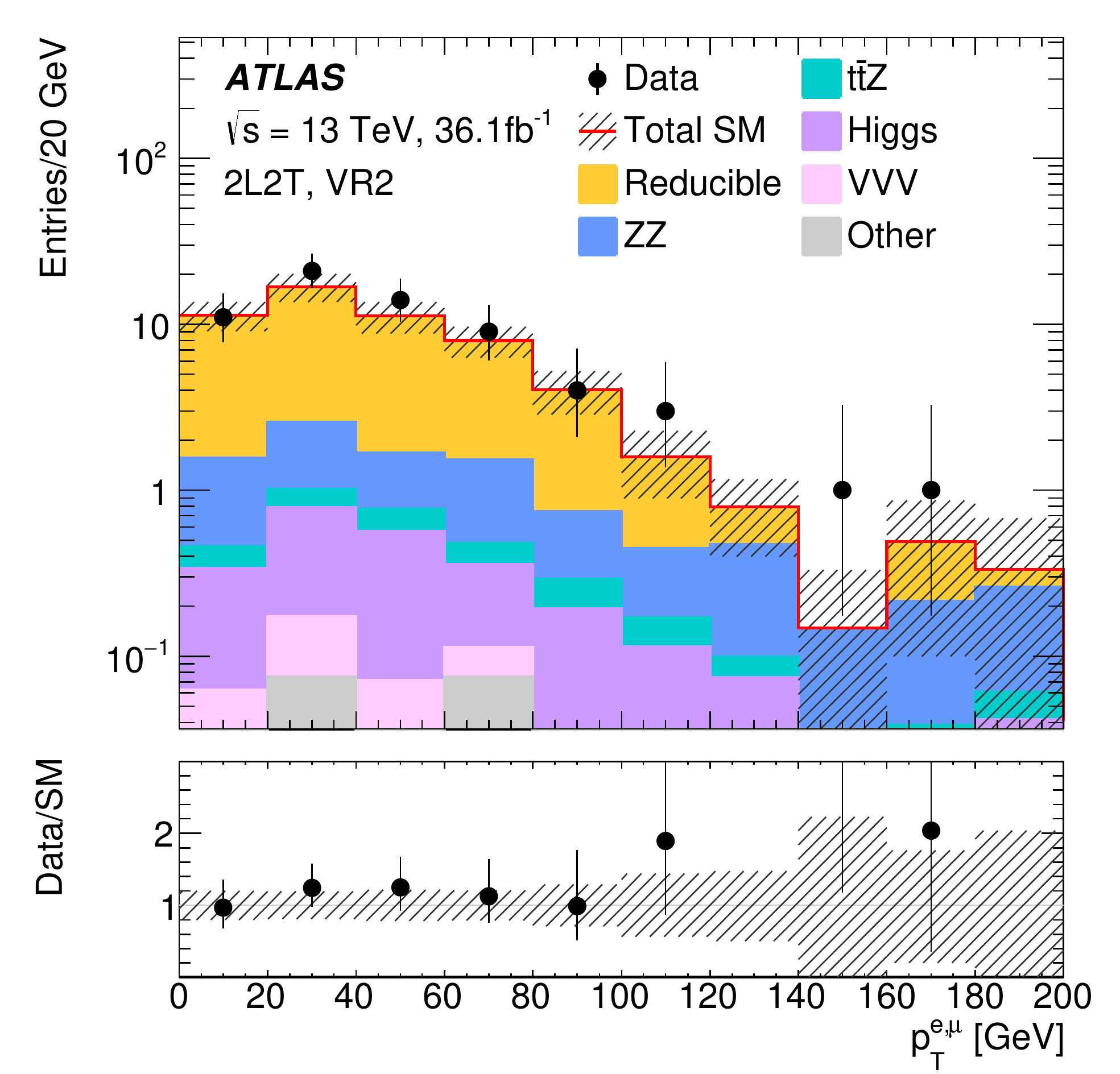}\label{fig:VR2LepPt}}
\subfloat[][$\pt\left({\tauhad}\right)$ in VR2]{\includegraphics[width=0.475\textwidth]{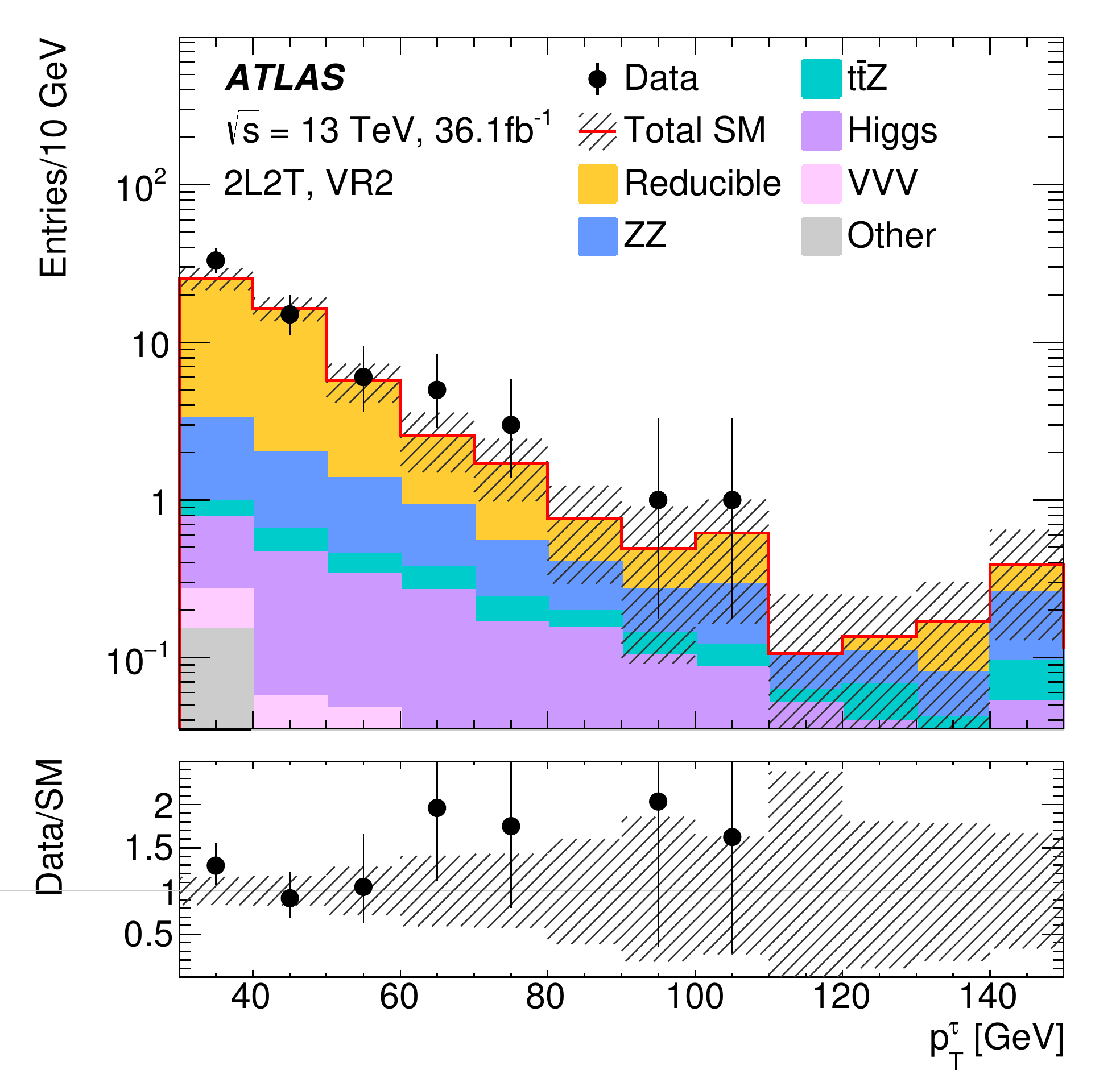}\label{fig:VR2TauPt}}
\caption{ The distributions for data and the estimated SM backgrounds in VR1 and VR2 for 
\protect\subref{fig:VR1LepPt} \&~\protect\subref{fig:VR2LepPt}~the light lepton \pt, and 
\protect\subref{fig:VR1TauPt} \&~\protect\subref{fig:VR2TauPt}~the tau \pt.
``Other'' is the sum of the \tWZ, \ttWW, and \tttt backgrounds.
The last bin includes the overflow. 
Both the statistical and systematic uncertainties in the SM background are included in the shaded band.
\label{fig:VR12LepPtAll}
}
\end{figure}

\FloatBarrier

\section{Results}\label{sec:results}

The expected and observed yields in each signal region are reported in Table~\ref{tab:SRyields}, together with the statistical and systematic uncertainties in the background predictions. 
The observations are consistent with the SM expectations within a local significance of at most $2.3\sigma$.
The \meff and \met distributions for all events passing signal region requirements, except the $\meff$ or $\met$ requirement itself, are shown in Figure~\ref{fig:SRmeffmet}.

\begin{table}[h]
\newcolumntype{R}{>{$}r<{$}}
\newcolumntype{L}{>{$}l<{$}}
\newcolumntype{M}{R@{${}\pm{}$}L}	
\centering
\small{
\begin{tabular}{ l MM MM MM } 
\toprule
Sample & \multicolumn{2}{c}{SR0A} & \multicolumn{2}{c}{SR0B}  & \multicolumn{2}{c}{SR0C} & \multicolumn{2}{c}{SR0D} & \multicolumn{2}{c}{SR1}  & \multicolumn{2}{c}{SR2}\\ \midrule
Observed & \multicolumn{2}{c}{$13$}& \multicolumn{2}{c}{$2$} & \multicolumn{2}{c}{$47$} & \multicolumn{2}{c}{$10$}  & \multicolumn{2}{c}{$8$} & \multicolumn{2}{c}{$2$} \\ \midrule
SM Total 	& $10.2 $&$ 2.1$      & $1.31 $&$ 0.24$      & $37 $&$ 9$   		& $4.1 $&$ 0.7$         & $4.9 $&$ 1.6$       & $2.3 $&$ 0.8$       \\ \midrule
$ZZ$  		& $2.7 $&$ 0.7$       & $0.33 $&$ 0.10$      & $28 $&$ 9$       	& $0.84 $&$ 0.34$       & $0.35 $&$ 0.09$     & $0.33 $&$ 0.08$     \\
$\ttbar Z$      & $2.5 $&$ 0.6$       & $0.47 $&$ 0.13$      & $3.2 $&$ 0.4$    	& $1.62 $&$ 0.23$       & $0.54 $&$ 0.11$     & $0.31 $&$ 0.08$     \\
Higgs 		& $1.2 $&$ 1.2$       & $0.13 $&$ 0.13$      & $0.9 $&$ 0.8$     	& $0.28 $&$ 0.25$       & $0.5 $&$ 0.5$       & $0.32 $&$ 0.32$     \\
$VVV$ 		& $0.79$&$ 0.17$      & $0.22 $&$ 0.05$      & $2.7 $&$ 0.6$       	& $0.64 $&$ 0.14$       & $0.18 $&$ 0.04$     & $0.20 $&$ 0.06$     \\
Reducible 	& $2.4 $&$ 1.4$       & \multicolumn{2}{c}{$0.000^{+0.005}_{-0.000}$} &\multicolumn{2}{c}{ $0.9^{+1.4}_{-0.9}$} & \multicolumn{2}{c}{$0.23^{+0.38}_{-0.23}$} & $3.1$&$ 1.5$    & $1.1 $&$ 0.7$  \\
Other 		& $0.53$&$ 0.06$      & $0.165 $&$ 0.018$    & $0.85 $&$ 0.19$         & $0.45 $&$ 0.10$       & $0.181$ & $0.022$    & $0.055 $&$ 0.012$     \\
\midrule
\multicolumn{1}{c}{$\langle\epsilon{\mathrm \sigma}\rangle_{\mathrm{obs}}^{95}~\mathrm{fb}$}
 & \multicolumn{2}{c}{$0.32$}
 & \multicolumn{2}{c}{$0.14$}
 & \multicolumn{2}{c}{$0.87$}              
 & \multicolumn{2}{c}{$0.36$}                 
 & \multicolumn{2}{c}{$0.28$}              
 & \multicolumn{2}{c}{$0.13$} \\[4pt]
\multicolumn{1}{c}{$S_{\mathrm{obs}}^{95}$}                                                           
& \multicolumn{2}{c}{$12$  }            
& \multicolumn{2}{c}{$4.9$}               
& \multicolumn{2}{c}{$31$}             
& \multicolumn{2}{c}{$13$}                 
& \multicolumn{2}{c}{$10$}              
& \multicolumn{2}{c}{$4.6$}   \\[4pt]
\multicolumn{1}{c}{$S_{\mathrm{exp}}^{95}$ }                                                         
& \multicolumn{2}{c}{$9.3^{+3.6}_{-2.3}$} 
& \multicolumn{2}{c}{$3.9^{+1.6}_{-0.8}$ }  
& \multicolumn{2}{c}{$23^{+8}_{-5}$}   
& \multicolumn{2}{c}{$6.1^{+2.1}_{-1.3}$}   
& \multicolumn{2}{c}{$6.5^{+3.5}_{-1.3}$} 
& \multicolumn{2}{c}{$4.7^{+2.0}_{-1.3}$} \\[4pt]
\multicolumn{1}{c}{$\mathrm{CL}_{\mathrm{b}}$ }      
& \multicolumn{2}{c}{$0.76$}             
& \multicolumn{2}{c}{$0.74$}            
& \multicolumn{2}{c}{$0.83$}            
& \multicolumn{2}{c}{$0.99$}                
& \multicolumn{2}{c}{$0.86$}             
& \multicolumn{2}{c}{$0.47$}\\
\multicolumn{1}{c}{$p_{0}$ }                                                                   
& \multicolumn{2}{c}{$0.23$ }          
& \multicolumn{2}{c}{$0.25$ }              
& \multicolumn{2}{c}{$0.15$}              
& \multicolumn{2}{c}{$0.011$}                 
& \multicolumn{2}{c}{$0.13$ }            
& \multicolumn{2}{c}{$0.61$} \\[4pt]
\multicolumn{1}{c}{$Z$}
& \multicolumn{2}{c}{$0.75$}              
& \multicolumn{2}{c}{$0.69$ }
& \multicolumn{2}{c}{$1.0$ }           
& \multicolumn{2}{c}{$2.3$ }              
& \multicolumn{2}{c}{$1.2$ }            
& \multicolumn{2}{c}{$0$}  \\
      \bottomrule
    \end{tabular} 
  }       
  \caption{Expected and observed yields for $36.1~\ifb$ in the signal regions. 
``Other'' is the sum of the \tWZ, \ttWW, and \tttt backgrounds.
Both the statistical and systematic uncertainties in the SM background are included in the uncertainties shown. 
Also shown are the model-independent limits calculated from the signal region observations;
the 95\% CL upper limit on
 the visible cross section times efficiency ($\langle\epsilon{\mathrm \sigma}\rangle_{\mathrm{obs}}^{95}$),
 the observed number of signal events($S^{95}_{\mathrm{obs}}$), and
 the signal events given the expected number of background events ($S^{95}_{\mathrm{exp}}$, $\pm1\sigma$ variations of the expected number) calculated by  performing  pseudo-experiments for each signal region. 
The last three rows report
 the CL$_{\mathrm{b}}$ value for the background-only hypothesis, and finally
 the one-sided $p_0$-value and the local significance $Z$ (the number of equivalent Gaussian standard deviations).
\label{tab:SRyields}}
\end{table}

\begin{figure}[h]
\centering
\subfloat[][$\meff$ in SR0A and SR0B]{\includegraphics[width=0.475\textwidth]{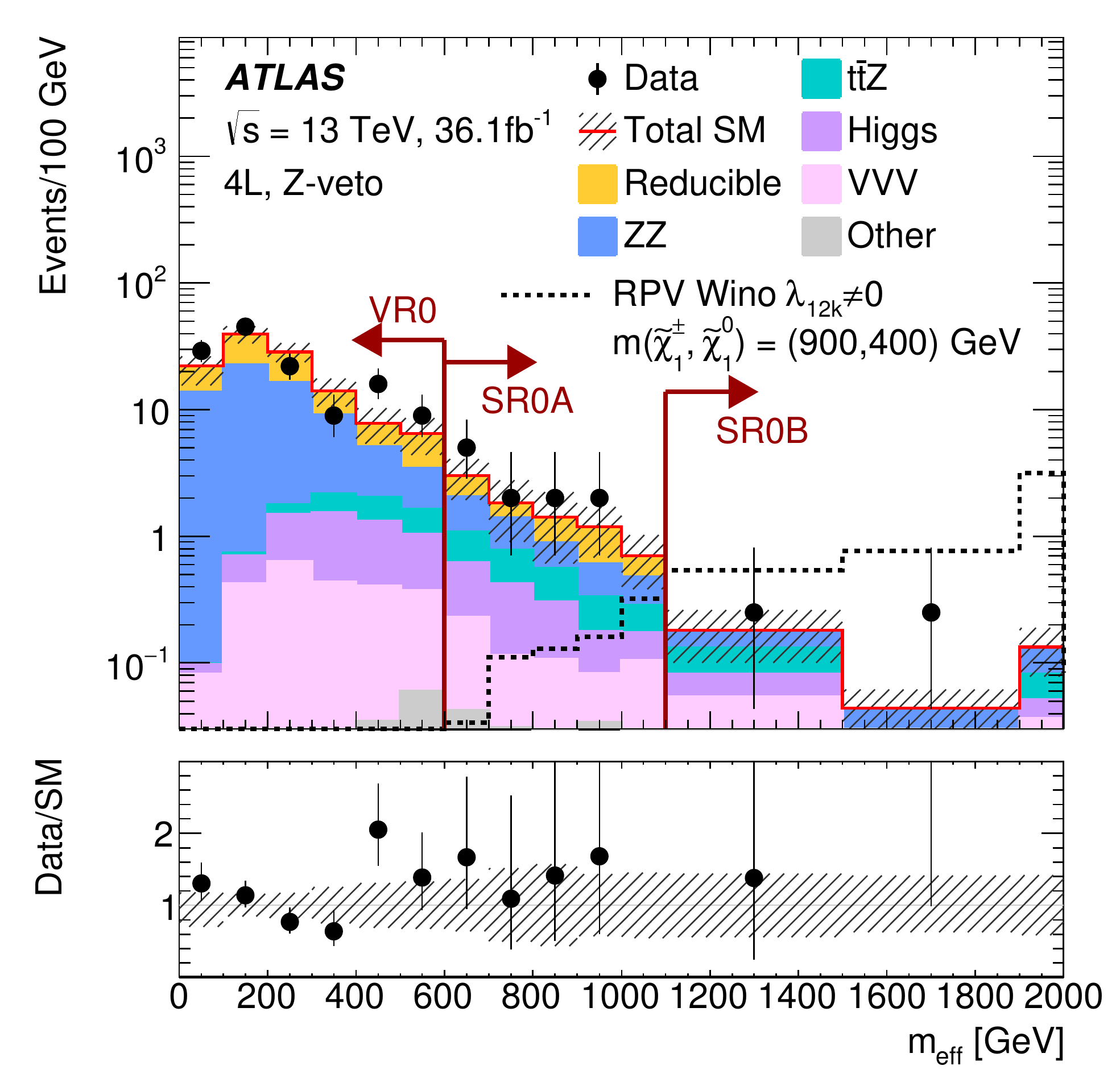}\label{fig:SR0noZmeff}}
\subfloat[][$\met$ in SR0C and SR0D]{\includegraphics[width=0.475\textwidth]{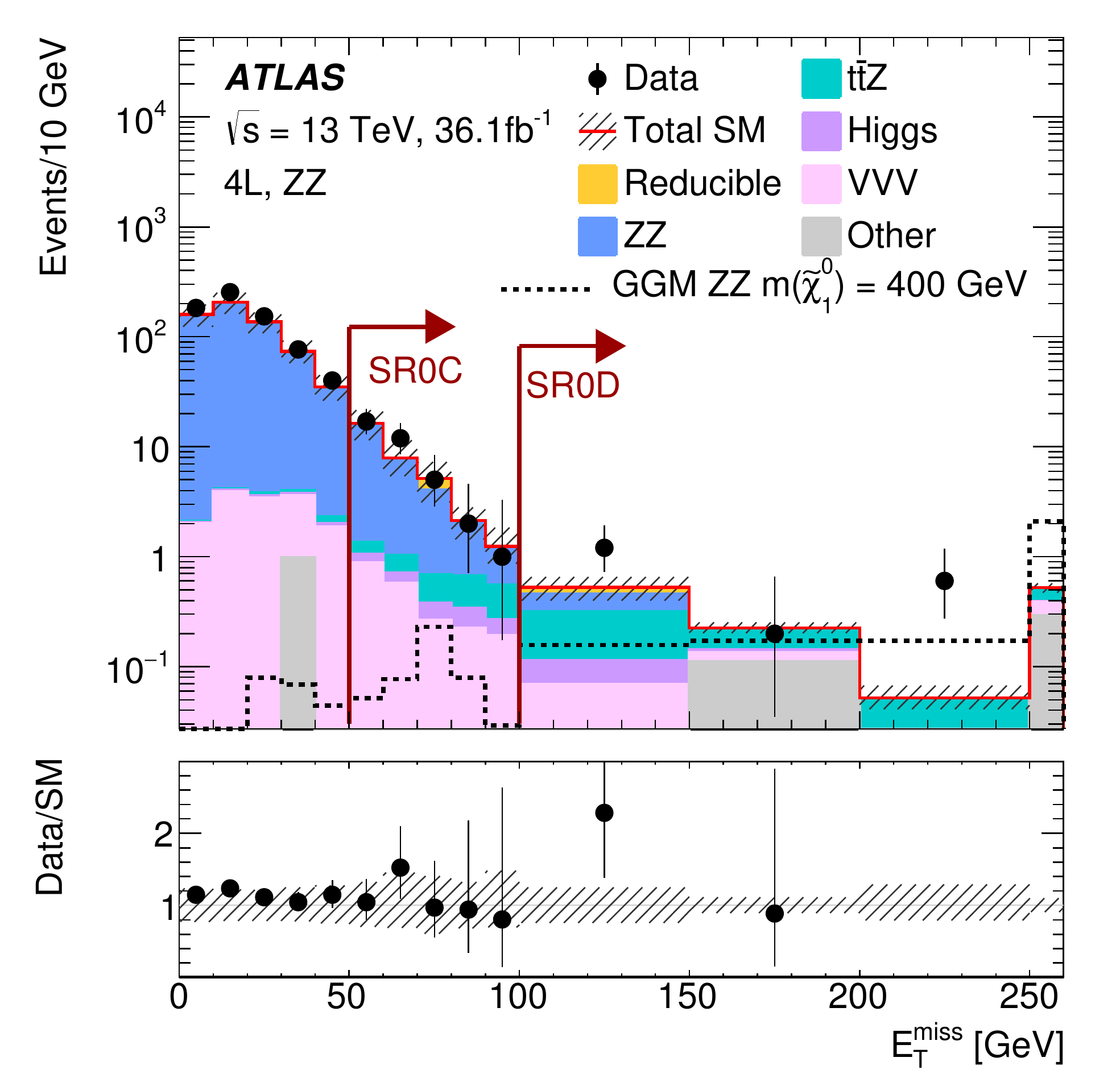}\label{fig:SR0Zmet}}\\
\subfloat[][$\meff$ in SR1]{\includegraphics[width=0.475\textwidth]{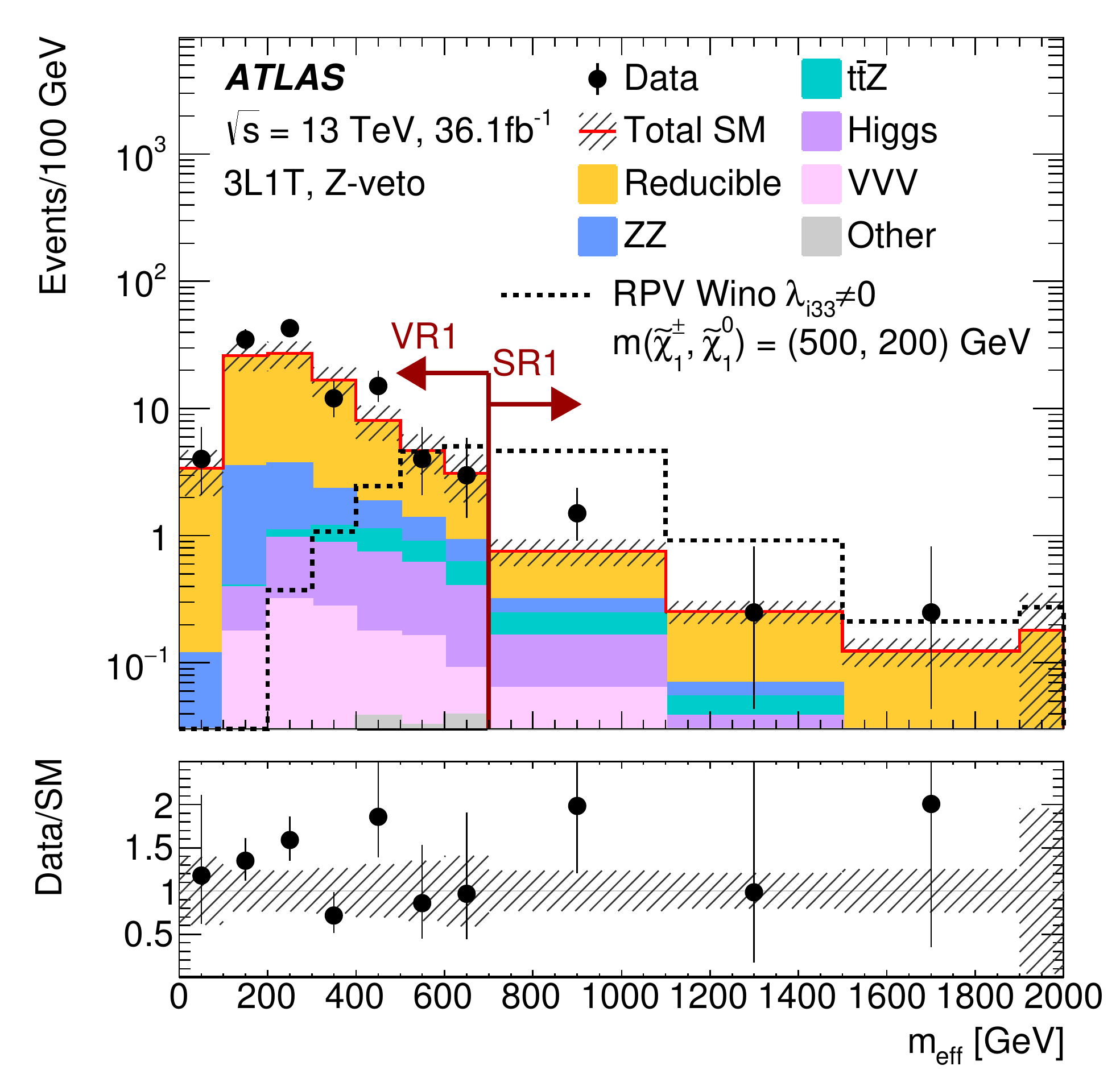}\label{fig:SR1meff}}
\subfloat[][$\meff$ in SR2]{\includegraphics[width=0.475\textwidth]{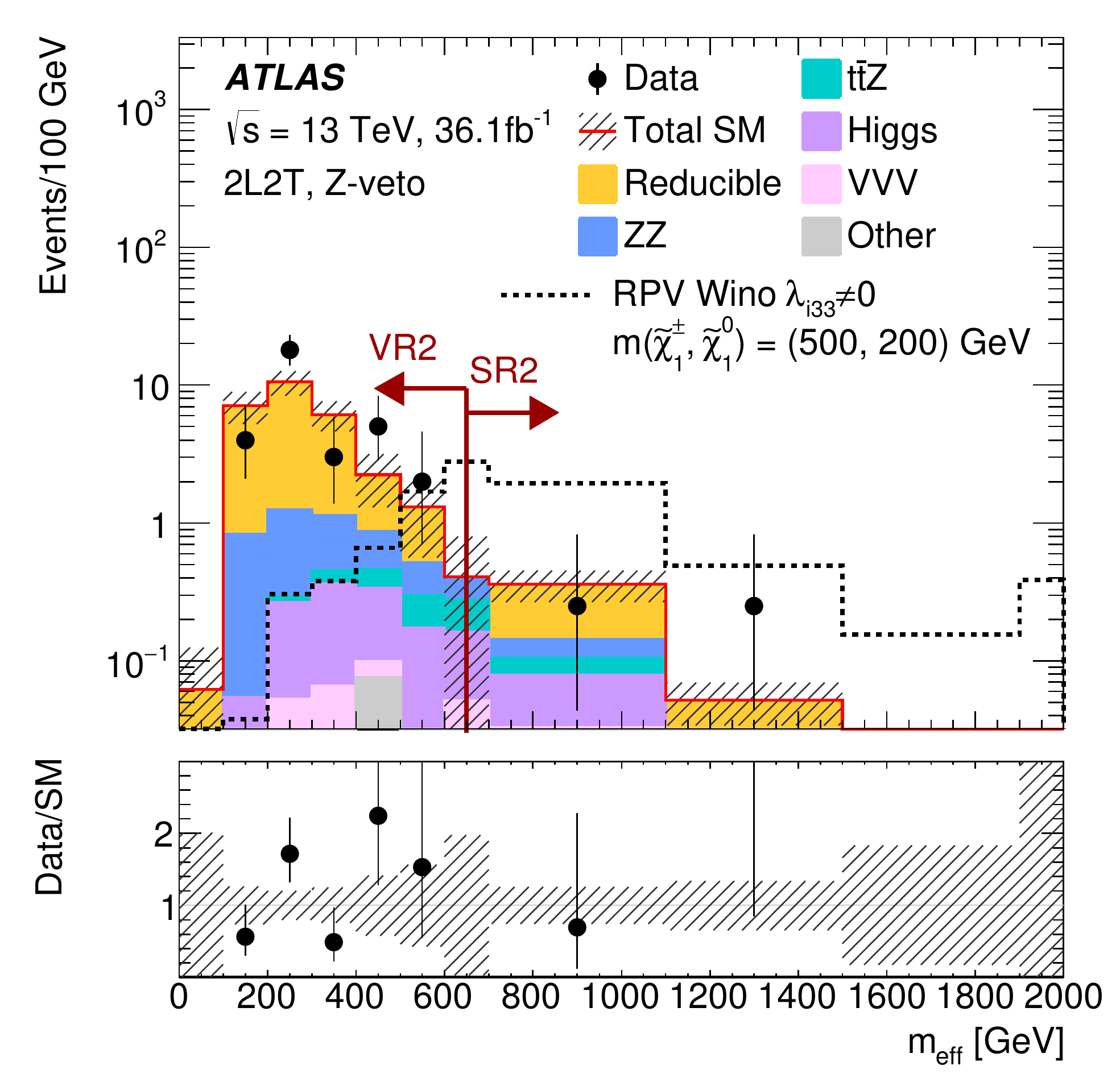}\label{fig:SR2meff}}
\caption{
The \protect\subref{fig:SR0noZmeff}, \protect\subref{fig:SR1meff} \& \protect\subref{fig:SR2meff} $\meff$ distribution
for events passing the signal region requirements except the $\meff$ requirement in SR0A, SR0B, SR1 and SR2.
The \protect\subref{fig:SR0Zmet}$~\met$ distribution is shown for events passing the signal region requirements except the $\met$ requirement in SR0C and SR0D. 
Distributions for data, the estimated SM backgrounds, and an example SUSY scenario are shown. 
``Other'' is the sum of the \tWZ, \ttWW, and \tttt backgrounds.
  The last bin captures the overflow events. 
Both the statistical and systematic uncertainties in the SM background are included in the shaded band. 
The red arrows indicate the $\meff$ or $\met$ selections in the signal regions.
\label{fig:SRmeffmet}
}
\end{figure}

The \HistFitter~\cite{Baak:2014wma} software framework is used for the statistical interpretation of the results. 
In order to quantify the probability for the background-only hypothesis to fluctuate to the observed number of events or higher, a one-sided $p_0$-value is calculated using pseudo-experiments, where the profile likelihood ratio is used as a test statistic~\cite{Cowan:2010js} to exclude the signal-plus-background hypothesis. 
A signal model can be excluded at 95\% confidence level (CL) if the CL$_{\mathrm{s}}$~\cite{Read:2002hq} of the signal-plus-background hypothesis is below $0.05$. 
For each signal region, the expected and observed upper limits at 95\% CL on the number of beyond-the-SM events ($S^{95}_{\mathrm{exp}}$ and $S^{95}_{\mathrm{obs}}$) are calculated using the model-independent signal fit.
The 95\% CL upper limits on the signal cross section times efficiency ($\langle\epsilon{\mathrm \sigma}\rangle_{\mathrm{obs}}^{95}$) and the CL$_{\mathrm{b}}$ value for the background-only hypothesis are also calculated for each signal region.

The number of observed events in each signal region is used to set exclusion limits in the SUSY models, where the statistical combination of all disjoint signal regions is used.
For overlapping signal regions, specifically SR0A and SR0B, and also SR0C and SR0D, the signal region with the better expected exclusion is used in the combination. 
Experimental uncertainties affecting irreducible backgrounds, as well as the simulation-based estimate of the weighted fake factors, are treated as correlated between regions and processes.
Uncertainties associated to the data-driven estimate of the reducible background are correlated between regions only. 
Theoretical uncertainties in the irreducible background and signal are treated as correlated between regions, while statistical uncertainties from MC simulation and data in the CR are treated as uncorrelated between regions and processes.  
For the exclusion limits, the observed and expected 95\% CL limits are calculated by performing pseudo-experiments for each SUSY model point, 
taking into account the theoretical and experimental uncertainties in the SM background and the experimental uncertainties in the signal. 
For all expected and observed exclusion limit contours, 
the $\pm1\sigma_\mathrm{exp}$ uncertainty band indicates the impact on the expected limit of the systematic and statistical uncertainties included in the fit. 
The $\pm1\sigma_\mathrm{theory}^\mathrm{SUSY}$ uncertainty lines around the observed limit illustrate the change in the observed limit as the nominal signal cross section is
scaled up and down by the theoretical cross section uncertainty.

Figure~\ref{fig:RPVexclLimits_1} and Figure~\ref{fig:RPVexclLimits_2} show the exclusion contours for the RPV models considered here, 
where SR0B dominates the exclusion for $\LLE 12k$ models, and the combination of SR1 and SR2 is important for the $\LLE i33$ models.
The exclusion limits in the RPV models extend to high masses, due to the high lepton multiplicity in these scenarios ($\ninoone\ra \ell\ell\nu$ with 100\% branching ratio) 
and the high efficiency of the $\meff$ selections.  
In the RPV wino NLSP $\LLE 12k$ models shown in Figures~\ref{fig:winoZbosonRPVexclLimits} and~\ref{fig:winoHiggsRPVexclLimit}, 
$\chinoonepm/\ninotwo$ masses up to $\sim 1.46\TeV$ are excluded for $m(\ninoone)>500\GeV$. 
The sensitivity is reduced for large mass splittings between the $\chinoonepm/\ninotwo$ and the $\ninoone$, where the decay products are strongly boosted, 
and $\chinoonepm/\ninotwo$ masses up to $\sim 1.32\TeV$ are excluded for $m(\ninoone)>50\GeV$.
Figures~\ref{fig:winoZbosonRPVexclLimits} and ~\ref{fig:winoHiggsRPVexclLimit} also show exclusion contours for the RPV wino NLSP $\LLE i33$ models, 
where $\chinoonepm/\ninotwo$ masses up to $\sim 980\GeV$ are excluded for $ 400\GeV < m(\ninoone) < 700\GeV$.
The sensitivity is also reduced for large mass differences between the $\chinoonepm/\ninotwo$ and the $\ninoone$, where the tau leptons, in particular, are collimated. 
These results extend the limits set in a similar model considering only $\chinoonep\chinoonem$ production in Ref.~\cite{SUSY-2013-13} by around $400$--$750 \GeV$.

Figure~\ref{fig:LHSlepRPVexclLimits} shows exclusion contours for the RPV $\sleptonL/\snu$ NLSP model, 
where left-handed slepton/sneutrino masses are excluded up to $\sim 1.06\TeV$ for $m(\ninoone)\simeq 600\GeV$ for $\LLE 12k$ models, 
and up to $780\GeV$ for $m(\ninoone)\simeq300\GeV$ for $\LLE i33$ models.
These results extend the limits set in a similar model considering only $\sleptonL\sleptonL$ production in Ref.~\cite{SUSY-2013-13} by around $200$--$400 \GeV$.

The exclusion contours for the RPV $\gluino$ NLSP model are shown in Figure~\ref{fig:gluinoRPVexclLimits},
where gluino masses are excluded up to $\sim 2.25\TeV$ for $m(\ninoone)>1\TeV$ for $\LLE 12k$ models, 
and up to $\sim 1.65\TeV$ for $m(\ninoone)>500\GeV$ for $\LLE i33$ models. 
These results significantly improve upon limits set in a similar model in Ref.~\cite{SUSY-2013-13} by around $500$--$700\GeV$.

Figure~\ref{fig:GGMexclLimits} shows the exclusion contours for the higgsino GGM models considered here.
The exclusion is dominated by SR0C and SR0D for low and high higgsino masses, respectively. 
Higgsino-like $\chinoonepm/\ninotwo/\ninoone$ with masses up to $295\GeV$ are excluded in scenarios with a branching ratio $\mathcal{B}(\ninoone \ra Z + \gravino) = 100\%$, 
while the exclusion is weakened for scenarios with $\mathcal{B}(\ninoone \ra Z + \gravino) < 100\%$.
This analysis is not sensitive to scenarios with $\mathcal{B}(\ninoone \ra h + \gravino) = 100\%$, 
where final states with lower lepton multiplicity may be more successful.  
The expected limit is comparable to those set using the combination of multiple analysis channels in Ref.~\cite{CMS-SUS-14-002}, 
but the observed limit is not as strong.

\begin{figure}[h]
\centering
\subfloat[][RPV wino \Wboson/\Zboson NLSP]{\includegraphics[width=0.8\textwidth]{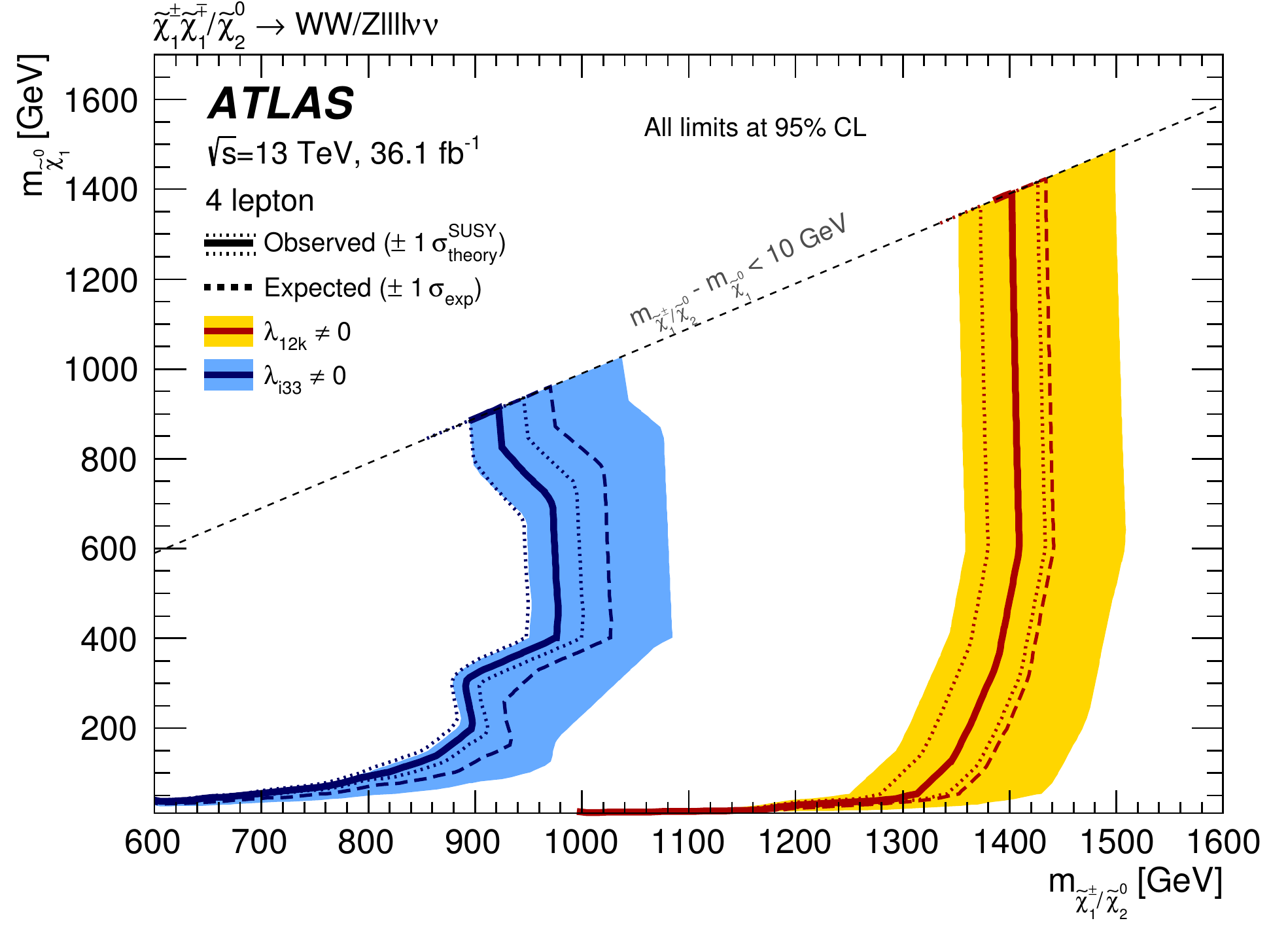}\label{fig:winoZbosonRPVexclLimits}}\\
\subfloat[][RPV wino \Wboson/\Higgs NLSP]{\includegraphics[width=0.8\textwidth]{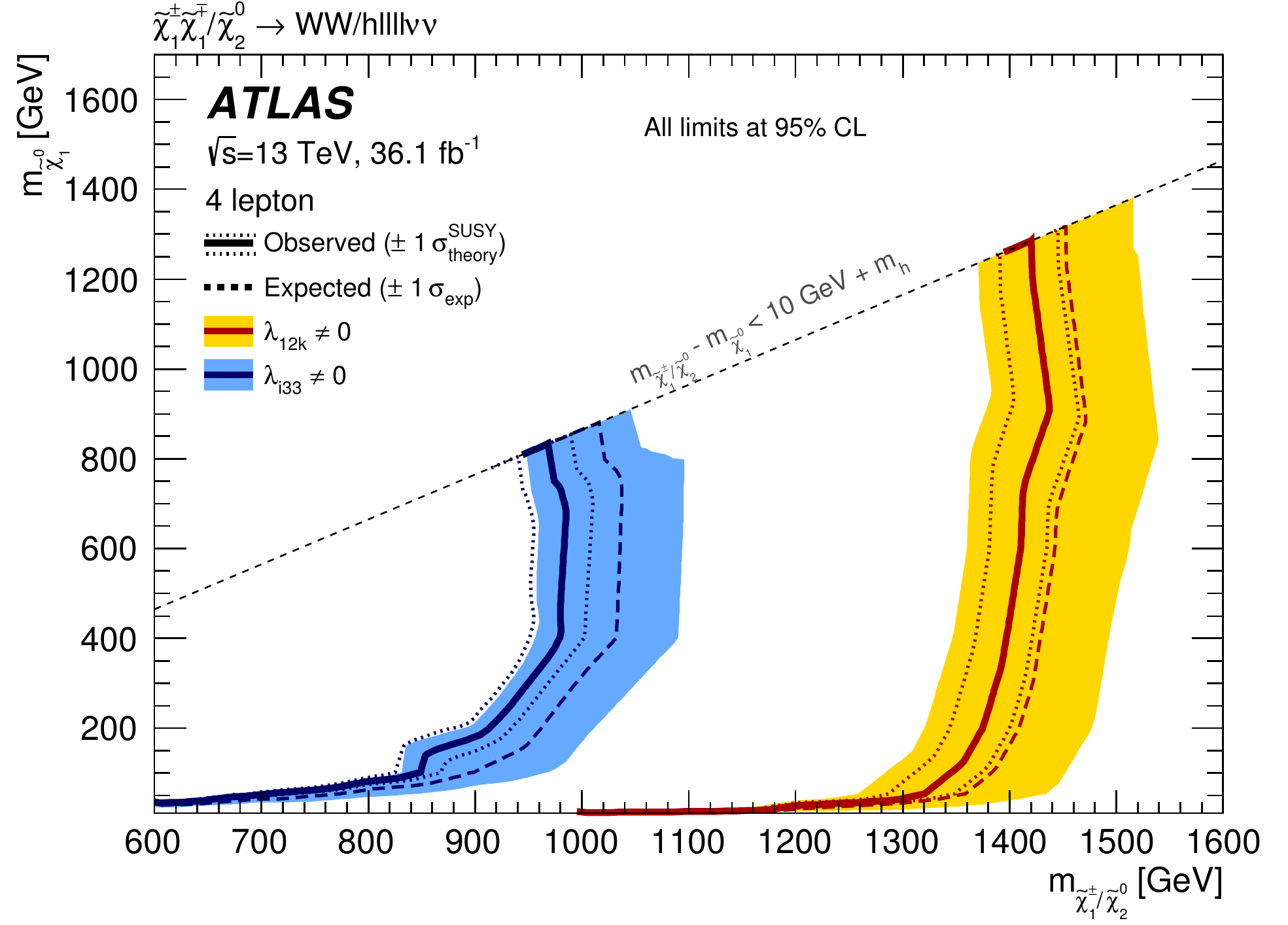}\label{fig:winoHiggsRPVexclLimit}} \\
\caption{
  Expected (dashed) and observed (solid) 95\% CL exclusion limits on
  \protect\subref{fig:winoZbosonRPVexclLimits}~wino \Wboson/\Zboson NLSP, and
  \protect\subref{fig:winoHiggsRPVexclLimit}~wino \Wboson/\Higgs NLSP pair production 
  with RPV $\ninoone$ decays via $\lambda_{12k}$, or $\lambda_{i33}$ where $i,k \in{1,2}$. 
The limits are set using the statistical combination of disjoint signal regions.
Where the signal regions are not mutually exclusive, the observed $\mathrm{CL}_{\mathrm{s}}$ value is taken from the signal region with the better expected $\mathrm{CL}_{\mathrm{s}}$ value.
\label{fig:RPVexclLimits_1}
}
\end{figure}

\begin{figure}[h]
\centering
\subfloat[][RPV $\sleptonL/\snu$ NLSP]{\includegraphics[width=0.8\textwidth]{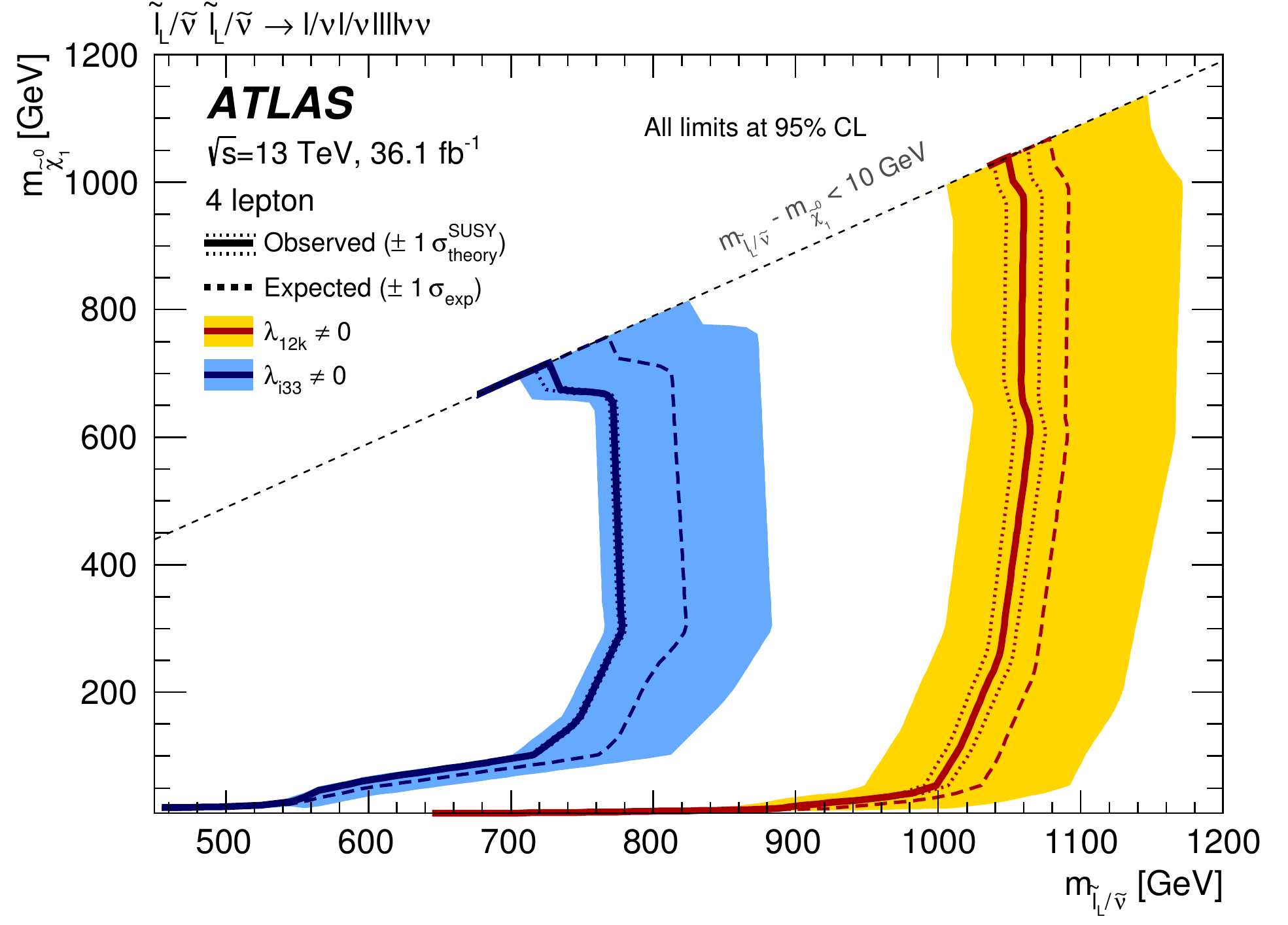}\label{fig:LHSlepRPVexclLimits}}\\
\subfloat[][RPV $\gluino$ NLSP]{\includegraphics[width=0.8\textwidth]{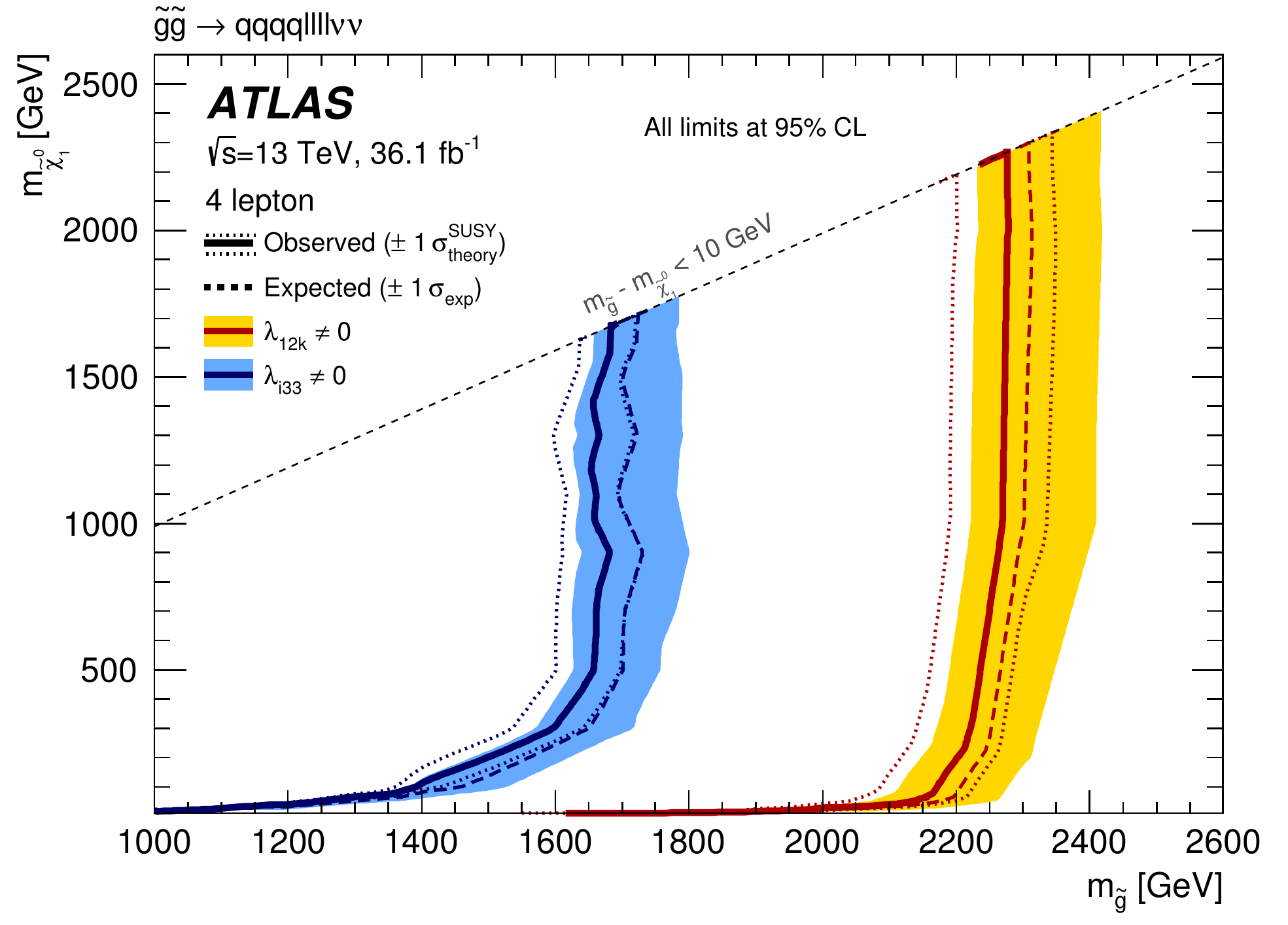}\label{fig:gluinoRPVexclLimits}}\\
\caption{
  Expected (dashed) and observed (solid) 95\% CL exclusion limits on
  \protect\subref{fig:LHSlepRPVexclLimits}~$\sleptonL/\snu$ NLSP, and
  \protect\subref{fig:gluinoRPVexclLimits}~gluino NLSP pair production 
  with RPV $\ninoone$ decays via $\lambda_{12k}$, or $\lambda_{i33}$ where $i,k \in{1,2}$.
The limits are set using the statistical combination of disjoint signal regions.
Where the signal regions are not mutually exclusive, the observed $\mathrm{CL}_{\mathrm{s}}$ value is taken from the signal region with the better expected $\mathrm{CL}_{\mathrm{s}}$ value.
\label{fig:RPVexclLimits_2}
}
\end{figure}

\begin{figure}[h]
\centering
\subfloat[][RPC GGM higgsino]{\includegraphics[width=0.8\textwidth]{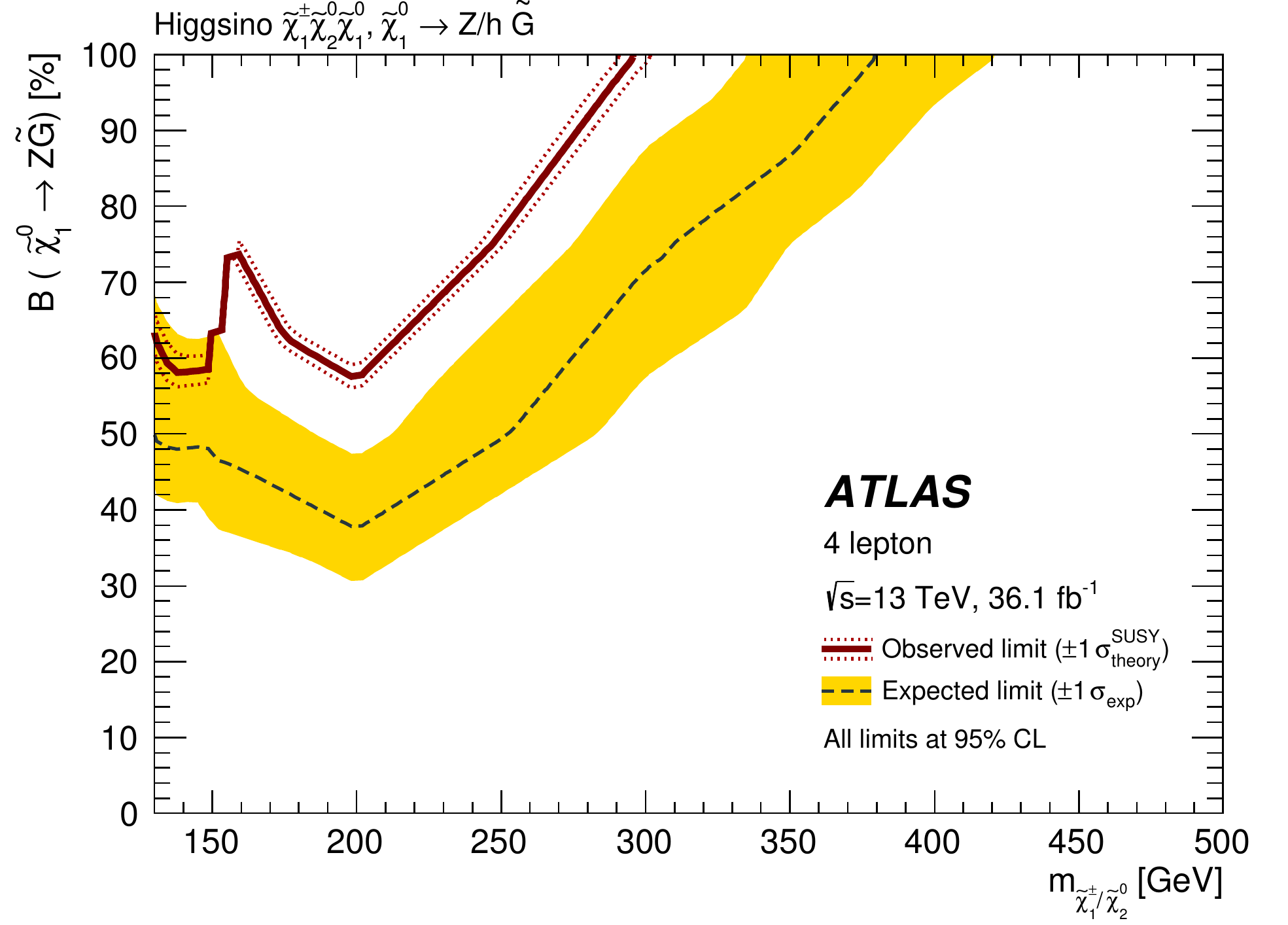}\label{fig:GGMexclLimits}} 
\caption{
  Expected (dashed) and observed (solid) 95\% CL exclusion limits on the higgsino GGM models.
The limits are set using the statistical combination of disjoint signal regions.
Where the signal regions are not mutually exclusive, the observed $\mathrm{CL}_{\mathrm{s}}$ value is taken from the signal region with the better expected $\mathrm{CL}_{\mathrm{s}}$ value.
\label{fig:GGMexclLimits}
}
\end{figure}

\FloatBarrier
\section{Conclusion}\label{sec:conc}

Results are reported from a search for new physics in the final state with four or more leptons (electrons, muons or taus), using $36.1\,\ifb$ 
of $\sqrt{s}=13\TeV$ proton--proton collision data collected by the ATLAS detector at the LHC in 2015 and 2016.
Six signal regions are defined with up to two hadronically decaying taus, and target lepton-rich RPV and RPC SUSY signals with selections requiring large effective mass or missing transverse momentum, and the presence or absence of reconstructed $Z$ boson candidates.
Data yields in the signal regions are consistent with Standard Model expectations. 
The results are interpreted in simplified models of NLSP pair production with RPV LSP decays, 
where wino-like $\chinoonepm/\ninotwo$, 
$\sleptonL/\snu$, 
and $\gluino$ masses up to 
$1.46\TeV$, 
$1.06\GeV$, 
and $2.25\TeV$ are excluded, respectively. 
The results are also interpreted in simplified higgsino GGM models, where higgsino-like $\chinoonepm/\ninotwo/\ninoone$ 
masses up to $295\GeV$ are excluded in scenarios with a 100\% branching ratio for $\ninoone$ decay to a $Z$ boson and a gravitino. 

\FloatBarrier
\section*{Acknowledgments}

We thank CERN for the very successful operation of the LHC, as well as the
support staff from our institutions without whom ATLAS could not be
operated efficiently.

We acknowledge the support of ANPCyT, Argentina; YerPhI, Armenia; ARC, Australia; BMWFW and FWF, Austria; ANAS, Azerbaijan; SSTC, Belarus; CNPq and FAPESP, Brazil; NSERC, NRC and CFI, Canada; CERN; CONICYT, Chile; CAS, MOST and NSFC, China; COLCIENCIAS, Colombia; MSMT CR, MPO CR and VSC CR, Czech Republic; DNRF and DNSRC, Denmark; IN2P3-CNRS, CEA-DRF/IRFU, France; SRNSFG, Georgia; BMBF, HGF, and MPG, Germany; GSRT, Greece; RGC, Hong Kong SAR, China; ISF, I-CORE and Benoziyo Center, Israel; INFN, Italy; MEXT and JSPS, Japan; CNRST, Morocco; NWO, Netherlands; RCN, Norway; MNiSW and NCN, Poland; FCT, Portugal; MNE/IFA, Romania; MES of Russia and NRC KI, Russian Federation; JINR; MESTD, Serbia; MSSR, Slovakia; ARRS and MIZ\v{S}, Slovenia; DST/NRF, South Africa; MINECO, Spain; SRC and Wallenberg Foundation, Sweden; SERI, SNSF and Cantons of Bern and Geneva, Switzerland; MOST, Taiwan; TAEK, Turkey; STFC, United Kingdom; DOE and NSF, United States of America. In addition, individual groups and members have received support from BCKDF, the Canada Council, CANARIE, CRC, Compute Canada, FQRNT, and the Ontario Innovation Trust, Canada; EPLANET, ERC, ERDF, FP7, Horizon 2020 and Marie Sk{\l}odowska-Curie Actions, European Union; Investissements d'Avenir Labex and Idex, ANR, R{\'e}gion Auvergne and Fondation Partager le Savoir, France; DFG and AvH Foundation, Germany; Herakleitos, Thales and Aristeia programmes co-financed by EU-ESF and the Greek NSRF; BSF, GIF and Minerva, Israel; BRF, Norway; CERCA Programme Generalitat de Catalunya, Generalitat Valenciana, Spain; the Royal Society and Leverhulme Trust, United Kingdom.

The crucial computing support from all WLCG partners is acknowledged gratefully, in particular from CERN, the ATLAS Tier-1 facilities at TRIUMF (Canada), NDGF (Denmark, Norway, Sweden), CC-IN2P3 (France), KIT/GridKA (Germany), INFN-CNAF (Italy), NL-T1 (Netherlands), PIC (Spain), ASGC (Taiwan), RAL (UK) and BNL (USA), the Tier-2 facilities worldwide and large non-WLCG resource providers. Major contributors of computing resources are listed in Ref.~\cite{ATL-GEN-PUB-2016-002}.


\FloatBarrier

\setcounter{maxnames}{5}
\printbibliography

\clearpage
 
\begin{flushleft}
{\Large The ATLAS Collaboration}

\bigskip

M.~Aaboud$^\textrm{\scriptsize 34d}$,    
G.~Aad$^\textrm{\scriptsize 99}$,    
B.~Abbott$^\textrm{\scriptsize 124}$,    
O.~Abdinov$^\textrm{\scriptsize 13,*}$,    
B.~Abeloos$^\textrm{\scriptsize 128}$,    
S.H.~Abidi$^\textrm{\scriptsize 165}$,    
O.S.~AbouZeid$^\textrm{\scriptsize 143}$,    
N.L.~Abraham$^\textrm{\scriptsize 153}$,    
H.~Abramowicz$^\textrm{\scriptsize 159}$,    
H.~Abreu$^\textrm{\scriptsize 158}$
% ...
\bigskip
\\

% ...
$^{13}$Institute of Physics, Azerbaijan Academy of Sciences, Baku; Azerbaijan.\\
% ...
$^{34}$$^{(a)}$Facult\'e des Sciences Ain Chock, R\'eseau Universitaire de Physique des Hautes Energies - Universit\'e Hassan II, Casablanca;$^{(b)}$Centre National de l'Energie des Sciences Techniques Nucleaires (CNESTEN), Rabat;$^{(c)}$Facult\'e des Sciences Semlalia, Universit\'e Cadi Ayyad, LPHEA-Marrakech;$^{(d)}$Facult\'e des Sciences, Universit\'e Mohamed Premier and LPTPM, Oujda;$^{(e)}$Facult\'e des sciences, Universit\'e Mohammed V, Rabat; Morocco.\\
% ...
$^{99}$CPPM, Aix-Marseille Universit\'e, CNRS/IN2P3, Marseille; France.\\
% ...
$^{124}$Homer L. Dodge Department of Physics and Astronomy, University of Oklahoma, Norman OK; United States of America.\\
% ...
$^{128}$LAL, Universit\'e Paris-Sud, CNRS/IN2P3, Universit\'e Paris-Saclay, Orsay; France.\\
% ...
$^{143}$Santa Cruz Institute for Particle Physics, University of California Santa Cruz, Santa Cruz CA; United States of America.\\
% ...
$^{153}$Department of Physics and Astronomy, University of Sussex, Brighton; United Kingdom.\\
% ...
$^{158}$Department of Physics, Technion, Israel Institute of Technology, Haifa; Israel.\\
$^{159}$Raymond and Beverly Sackler School of Physics and Astronomy, Tel Aviv University, Tel Aviv; Israel.\\
% ...
$^{165}$Department of Physics, University of Toronto, Toronto ON; Canada.\\
% ...
$^{a}$ Also at  Department of Physics, University of Malaya, Kuala Lumpur; Malaysia.\\
$^{b}$ Also at Borough of Manhattan Community College, City University of New York, NY; United States of America.\\
$^{c}$ Also at Centre for High Performance Computing, CSIR Campus, Rosebank, Cape Town; South Africa.\\
$^{d}$ Also at CERN, Geneva; Switzerland.\\
$^{e}$ Also at CPPM, Aix-Marseille Universit\'e, CNRS/IN2P3, Marseille; France.\\
% ...
$^{*}$ Deceased

\end{flushleft}


\end{document}